\documentclass[12pt]{article}
\usepackage{amsmath, amssymb, graphics}
\usepackage{setspace}
\usepackage[note]{achemso}
\usepackage{amsmath}
\usepackage{multicol}
\usepackage{subfigure}
\usepackage{graphicx}
\usepackage{rotating}
\usepackage{geometry}
\usepackage{verbatim}
\usepackage{multirow}
\usepackage{comment}
\usepackage{amssymb}
\usepackage{bm}
\newcommand{\mathsym}[1]{{}}

\begin{document}

\title{Minimal models for proteins and RNA: From folding to function}
\author{David L. Pincus$^{1}$, Samuel S. Cho$^{1}$, Changbong Hyeon$^3$ and D. Thirumalai$^{1,2,}$\footnote{Corresponding author phone: 301-405-4803; fax: 301-314-9404; thirum@umd.edu}}
\maketitle
\noindent $^1$Biophysics Program, Institute for Physical Science and Technology\\
$^2$Department of Chemistry and Biochemistry\\
University of Maryland, College Park, MD 20742\\
$^3$Department of Chemistry, Chung-Ang University, Seoul 156-756, Republic of Korea

\begin{abstract}
We present a panoramic view of the utility of coarse-grained (CG)
models to study folding and functions of proteins and RNA. Drawing
largely on the methods developed in our group over the last twenty
years, we describe a number of key applications ranging from folding of proteins with
disulfide bonds to functions of molecular machines. After presenting the
theoretical basis that justifies the use of CG models, we explore the
biophysical basis for the emergence of a finite number of folds from 
lattice models.  The lattice model simulations of approach to the folded state show that non-native
interactions are relevant only early in the folding process - a
finding that rationalizes the success of structure-based models that
emphasize native interactions.  Applications of off-lattice
$C_{\alpha}$ and models that explicitly consider side chains ($C_{\alpha}$-SCM) to
folding of $\beta$-hairpin and effects of macromolecular crowding are 
briefly discussed.  Successful applications of a new class of off-lattice
models, referred to as the Self-Organized Polymer (SOP), intended 
to probe dynamics in large proteins is illustrated by describing
the response of Green Fluorescent Protein (GFP) to mechanical force.
The unfolding of GFP, at constant loading rate, occurs by the kinetic
partitioning mechanism, involving a bifurcation in the pathways. The
utility of the SOP model is further illustrated by applications that clarify the 
functions of the chaperonin GroEL and motion of the molecular motor
kinesin.  We also present two distinct models for RNA, namely, the Three Site
Interaction (TIS) model and the SOP model, that probe forced
unfolding and force quench refolding of a simple hairpin and {\it
Azoarcus} ribozyme. The unfolding pathways of {\it Azoarcus} ribozyme depend on the 
loading rate, while constant force and constant loading rate
simulations of the hairpin show that both forced-unfolding and
force-quench refolding pathways are heterogeneous.  The location of the transition state 
moves as force is varied.  The predictions based on the SOP model show that
force-induced unfolding pathways of the ribozyme can be dramatically
changed by varying the loading rate.  We conclude with a discussion 
of future prospects for the use of coarse-grained models in addressing problems of outstanding 
interest in biology.
\end{abstract}

\section*{Introduction}

In order to carry out the myriad of cellular functions proteins \cite{Creighton:1984,FershtBook} and RNA \cite{DoudnaNature02} have to fold to well defined three dimensional structures.
Protein folding is a process by which a polypeptide chain made up of a linear sequence of amino acids self-assembles into a compact
three dimensional structure.   Experiments show that single domain proteins reach their native states on the time scales on the order of
10-1000 milliseconds \cite{EatonARBBS00}, which is rapid given the potential complexity of the folding process.  Besides the intellectual challenge, solution of the protein folding
problem will have important applications in the design of enzymes that can carry out non-biological reactions and in biotechnology.  Moreover, the quest to understand 
how proteins fold has become important because 
misfolding and subsequent aggregation of proteins has been linked to a number of diseases (Alzheimer's disease, prion disorders, CJD, Parkinsons are few of
the more common ones known to date) \cite{ChitiARB06,ThirumalaiCOSB03,TreiberCOSB99,Dobson99TBS,SelkoeNature03}. In the last two decades, considerable progress has been made in attaining a global understanding
of the mechanisms by which proteins fold thanks to breakthroughs in experiments \cite{FershtCell02,SchulerCOSB08,Jackson98FD}, theory \cite{OnuchicCOSB04,HyeonBC05,ShakhnovichChemRev06}, 
and computations\cite{KuhlmanScience03, SnowAnnuRev05, DobsonAngewante05, SheaARPC01}.  Fast folding experiments \cite {EatonPNAS93,KiefhaberPNAS99,EatonARBBS00,SchulerCOSB08, DenizPNAS00} and single molecule methods \cite{SosnickCOSB03, Haran03PNAS, FernandezSCI04}
have begun to provide a direct glimpse into the initial stages of protein folding.  
These experiments show that there is a great diversity in the routes explored during the  transitions from unfolded states 
to the folded state that were unanticipated in ensemble experiments.  In particular, the use of mechanical force to generate folding trajectories shows that
the pathways explored in the folding process can vary greatly depending on the initial location in the folding landscape 
from which folding is commenced \cite{FernandezSCI04}.  The advantage of single molecule experiments, 
which use force to initiate folding, is that they can explore regions of the energy landscape that are totally inaccessible in conventional methods in which folding processes are probed by
changing denaturant concentration or temperature \cite{FershtBook}.  These increasingly sophisticated experiments have ushered in an era in which new theoretical models are needed to make quantitative and testable predictions.

In contrast to the intense effort in deciphering the folding mechanism of proteins, the study of the self-assembly of RNA molecules began in earnest only after the landmark discovery
that RNA can also perform catalytic activity \cite{CechCell81,CechCell82,AltmanCell83,AltmanSci84}. In the intervening years, an increasing repertoire of cellular functions have been associated with RNA \cite{DoudnaNature02}. 
These include their role in replication, translational regulation, and viral propagation. 
Moreover, interactions of RNA with each other and with DNA and proteins are vital in many biological processes.  
Furthermore, the central chemical activity of ribosomes, namely, the formation of the peptide bond in the biosynthesis of polypeptide chains near the peptidyl transfer center, involves only RNA leading many to suggest that ribosomes are ribozymes \cite{SteitzSCI00,NollerSCI01,YonathNature01}. 
The appreciation that RNA molecules play a major role in a number of cellular functions has made it important to establish their structure-function relationships. 
Just as in the case of proteins, the last fifteen years have also witnessed great strides in dissecting the complexity of RNA folding\cite{TreiberCOSB01, ThirumalaiACR96, SosnickCOSB03}.  
The number of experimentally determined high resolution RNA structures \cite{CateStructure96,SteitzSCI00,NollerSCI01} continues to increase, enabling us to understand the interactions that stabilize the folded states. 
Single molecule \cite{Bustamante03Science,Tinoco06PNAS,RussellJMB01,WoodsideScience06,ZhuangSCI00,Ma06JACS} and ensemble experiments \cite{ZarrinkarNSB94,KoculiJMB06,PanJMB99} using a variety of biophysical methods combined with theoretical techniques \cite{ThirumalaiACR96,HyeonBC05} have led to a conceptual framework for predicting various mechanisms by which RNA molecules fold. 
In order to make further progress new computational tools are required.  
Simulations of RNA molecules are difficult because their folding invariably requires counterions.  Accounting for electrostatic interactions, which operate on multiple length scales, 
is a notoriously difficult problem.  Nevertheless, as we document here, the principles that justify the use of minimal models for proteins can also be used to model RNA.

Because functions of ribozymes and proteins are linked to folding, that may occur either spontaneously or in association with other biomolecules, we are inevitably lead to the question: How do these molecules fold?  
In this review, we describe insights into the folding mechanisms of proteins and RNA that have come from using coarse grained (CG) models. 
In principle, many of the important questions in biomolecular folding and their functions can be addressed using all-atom Molecular Dynamics (MD) simulations in explicit water\cite{SnowAnnuRev05}. 
While this approach is valuable in many contexts, it is difficult to simulate the processes of interest described in this article reliably for long enough times to obtain insights or make testable predictions. 
As a result, there has been a great emphasis on developing CG models that capture the essential physics of the processes of interest.  
The major advantage of CG models, many of which were developed in our group over the past twenty years, is that accurate simulations can be carried out.  
The CG models have been of great importance in explaining a number of experimental observations, and they have also led to several successful predictions. Indeed, as the system size gets larger, as is the case for molecular machines for example, a straightforward MD approach cannot currently be used to follow the complex conformational changes the enzymes undergo during their reaction cycle.  
The use of CG models is not merely a convenience. Indeed, as we argue in the next section, there is a theoretical basis for using the structure-based models for folding and function.  
Here, we show using largely problems that we have solved, that simulations of CG model for complex problems accompanied by theoretical arguments 
have become the mainstay in addressing some of the outstanding issues in the folding and function of proteins and RNA.

\subsection*{Rationale for developing Structure-Based CG models}

The use of coarse-grained models has a rich history in physics. 
In particular, models that capture the essence of a phenomena have been crucial in condensed matter physics \cite{Anderson:1997} and soft matter science \cite{deGennesbook} - areas that are most closely related to the subject matter of the present article.  
For example, it is well known that spin systems are excellent models for a quantitative understanding of magnetism.  
Similarly, the complex phenomenon of superconductivity can be understood without accounting for all of the atomic details of the constituent matter\cite{ZhangBiochem93,deGennesSCMA}.  
In polymer physics, several universal properties, such as the dependence of the size, $R_g$, of the polymer on the number of monomers, as well as the distribution of the end-to-end distances, only depend on the solvent quality and not on the details of the monomer structure \cite{Yamakawabook,Florybook,deGennesbook}  There are firm theoretical bases for using minimal models to describe complex phenomena such as those highlighted above.  The concept of universality, embedded in the theory of critical phenomena \cite{Ma:1976} and expressed in renormalization group theory \cite{WilsonRMP83}, assures us that near the critical point the system is dominated by only one dominant (divergent) length scale.  Hence, the universal properties, such as the vanishing of the order parameter or the divergence of specific heat, depend only on dimensionality-determined critical exponents.  Similarly, the mapping of the problem of a polymer in a good solvent (also referred to as Òself-avoiding walkÓ) to an $n$-vector spin model with $n \rightarrow 0$ established a firm link between the universal behavior of polymers and critical phenomenon\cite{LevittNature75}, thereby explaining the Flory law for the variation of $R_g$ as a function of $N$, the number of monomers \cite{desCloizeaux75JP,deGennes72PhysLett,deGennes75Macro}.  More importantly, such a mapping showed why the critical exponents, known in magnetic models, arise in the description of polymer properties, regardless of the chemical details of the monomers. 

In the context of biopolymers, phenomenological theories have helped rationalize the use of CG models.  Although such theories are not as sound as the ones alluded to in the previous paragraph, they do take into account evolutionary considerations that are difficult to model with the same rigor as some of the phenomena in the physical and material world.  The realization that evolved biopolymers such as RNA and proteins must be different came from theoretical studies  of random heteropolymer and related models\cite{BryngelsonJCP89, ShakhnovichBiophysChem1989, Garel96}.  These studies showed that proteins made of random sequences cannot kinetically access the unique functional states on biologically relevant time scales.  In particular, the dynamics of these models showed that typically random sequences would be stuck in metastable states for arbitrary long times,  thus displaying glass-like  behavior\cite{ThirumalaiPRL96, TakadaPNAS97}. From these studies, it followed that the evolutionary process has resulted in proteins and RNA sequences   that can fold and be (marginally) stable during their cellular life cycle.  These ideas, that distinguish evolved proteins and those that are generated from random sequences, can be cast more precisely in terms of the characteristic temperatures that describe the potential conformational transitions in proteins. The temperatures that control foldability (efficient folding without being kinetically trapped in the competing basins of attraction (CBAs) for times so long that aberrant processes like aggregation become relevant) are the collapse temperature, $T_{\theta}$\cite{CamachoPNAS93}, the folding transition temperature $T_F$, and the glass transition temperature $T_g$\cite{GoldsteinPNAS92}.  At the temperature $T_{\theta}$ (named in honor of Flory), proteins collapse into compact structures from an expanded coil, and at $T = T_F$ they undergo a transition to the folded native state. The relaxation dynamics at the glass transition temperature $T_g$ slows down the conformational changes to a great extent, thus resulting in kinetic trapping in a large number of metastable minima\cite{SocciJCP95}.  Theoretical considerations were used to show that in foldable sequences $T_g < T_F$\cite{GoldsteinPNAS92,GarelJPII94}.  Alternatively, it was suggested that the avoidance of trapping in deep CBA's for long times requires that $T_F \approx T_{\theta}$\cite{CamachoPNAS93}.  Indeed, it was shown based on the treatment of dynamics of heteropolymer models\cite{ThirumalaiPRL96} and simple arguments that the two criteria are, in all likelihood, related.  Using explicit calculations on a random hydrophobic-hydrophilic model\cite{ThirumalaiPRL96} Thirumalai, Ashwin, and Bhattacharjee showed that 
\begin{equation}
\label{eqn:ThetaOverG}
T_\theta / T_g = \frac{\sqrt{1+40\beta N} + 1}{2}
\end{equation}
\noindent It follows from Eq. (\ref{eqn:ThetaOverG}) that for a given $N$, ($T_\theta/T_g$) increases as the ratio ($\beta$) between the three and two body interaction strength increases.  For
$T_\theta/T_g \approx 6$, which coincides with the value for $T_F/T_g$ proposed by Kaya and Chan \cite{KayaProteins00}, we get from Eq. (\ref{eqn:ThetaOverG}) $\beta = 3/N$.  Thus, for proteins in the size range
corresponding to protein L $\beta \approx 0.05 $ which shows that modest three-body interaction suffices to maximize $T_\theta/T_g$, and hence $T_F/T_g$ because $\max(T_F/T_g) \approx T_\theta/T_g$.  We should emphasize that $T_g$ in Eq.\ (\ref{eqn:ThetaOverG}) is
a kinetic glass transition temperature and not the thermodynamic temperature at which conformational entropy vanishes.  It is important to realize that the characteristic temperatures that describe foldable sequences depend on the entire free energy `spectrum' of protein conformations, which implies that the entropy of the misfolded states have to be included in the calculation of $T_F$, $T_{\theta}$, and $T_g$\cite{KlimovJCP98}. 

What is the connection between inequalities relating the characteristic temperatures ($T_g < T_F \le T_\theta$) and models of proteins that exhibit protein-like behavior? It has been suggested that the energy landscape of foldable sequences is smooth and `funnel'-shaped so that they can be navigated efficiently \cite{Bryngelson95Protein,LeopoldPNAS92}.  We interpret funnel-shaped to mean that the gradient of the large dimensional energy landscape towards the native basin of attraction (NBA) is `large' enough that the biomolecule does not get kinetically trapped  in  the CBAs for long times during the folding process. However, sequences with perfectly smooth energy landscapes are difficult to realize because of energetic and topological 
frustration \cite{ThirumalaiACR96,ClementiJMB00}. 
In proteins, the hydrophobic residues prefer to be sequestered in the interior 
while polar and charged residues are better accommodated on the surfaces 
where they can interact with water. 
Often these conflicting requirements cannot be simultaneously satisfied, and hence proteins and RNA can be energetically ``frustrated''. 
In all likelihood, only evolved or well designed sequences can minimize 
energetic frustration. Even if a particular foldable sequence minimizes energetic conflicts, it is 
nearly impossible to eliminate topological frustration, especially in large proteins, which arises due to 
chain connectivity \cite{GuoBP95,ThirumalaiRNA00}. 
If the packing of locally formed structures is in conflict with the global
 fold then the polypeptide or polynucleotide chain is topologically frustrated\cite{KiefhaberPNAS1995}. 
Both sources of frustration, energetic and topological, render the energy landscape rugged 
on length scales that are larger than those in which secondary structures ($\approx (1-2)$ nm) form even if folding can be globally described using only two-states (i.e., folded and unfolded). These conflicting demands are minimized for sequences with a large gradient towards the native basins of attraction (NBA's).

An immediate and crucial consequence of realizing that energetic frustration is minimized in natural proteins is that the strength of the interactions between amino acid residues that are present in the native state characterized by a free energy scale $g _N$ must be stronger than the non-native (i.e., those that are not in the native state) interactions ($g_{NN}$).  The inequality $g _N/g _{NN}$ assures
us that the NBA is kinetically accessible under a wide range of external conditions.  For the well designed sequences, non-native interactions play a role only in the 
initial stages of the folding process as early simulations using lattice models showed (see below).  Fluctuation in $g _{NN}$ and the associated entropy of the non-native conformations (ones whose overlap with the native state is substantial) is related to the characteristic temperatures\cite{KlimovJCP98}.  If $g _N/g _{NN} > 1$, then the gradient towards the native conformation from any part of the folding landscape would be greater than the spatial variation in the underlying roughness, which we view as a mathematical definition of the funnel-shaped landscapes. From these arguments, it follows that CG models that emphasize the role of native over non-native interactions can provide a valuable description of folding, assembly, and function(s) of biomolecules.  The success of such an approach in a wide range of applications is a testimony to the use of CG models, and the underlying conceptual basis for their use.    

\section*{Protein Folding}

\subsection*{Lattice Models} 
\begin{figure}[ht]
\includegraphics[width=7.00in]{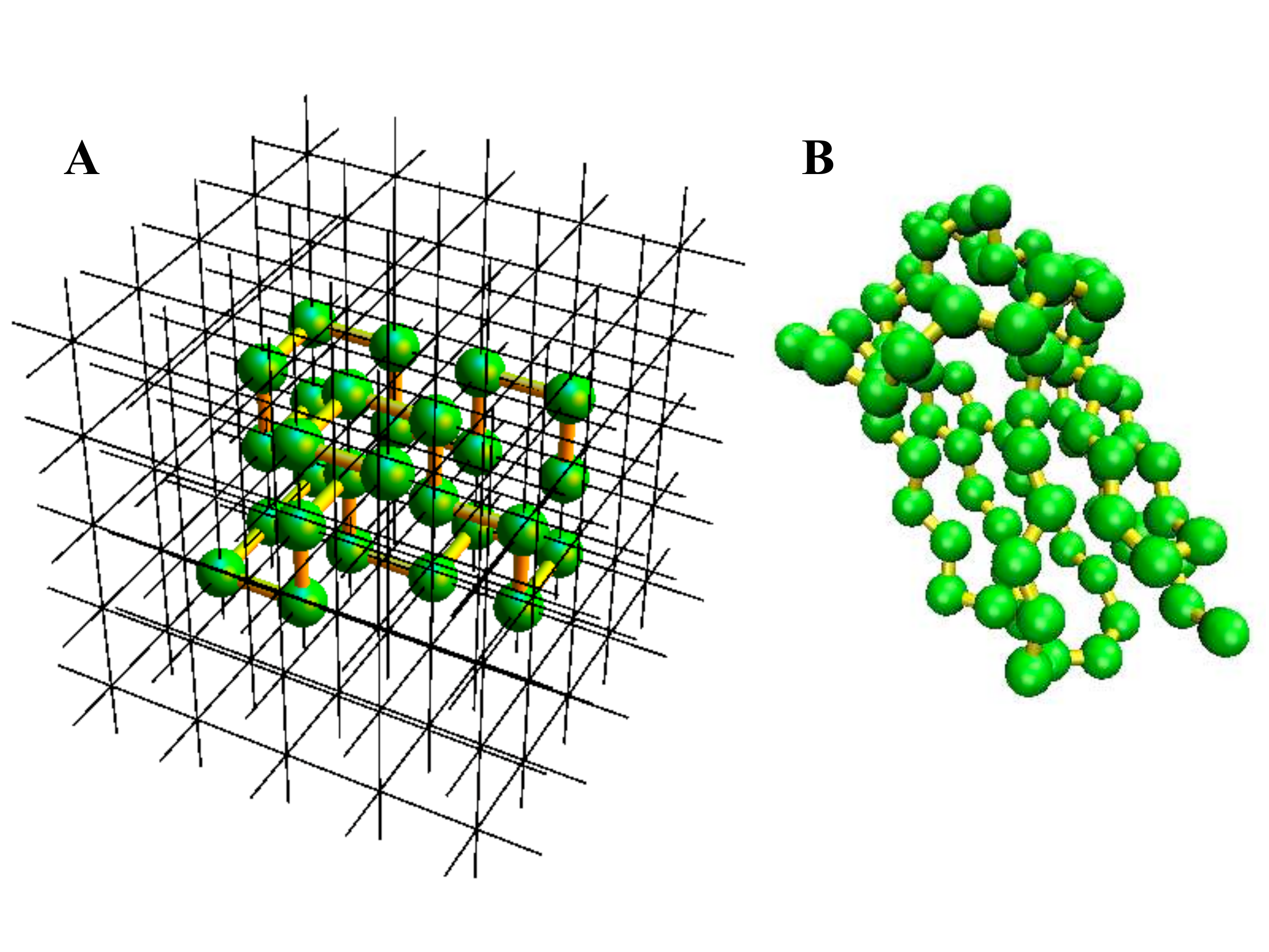}
\caption{}
\label{fig:1}
\end{figure}

Lattice models were used in the early 1950s to calculate the universal properties of self-avoiding random walks.  A familiar approach was advocated in the late eighties to study protein folding\cite{LauMacro89, ChanJCP89}.  In applications to proteins, two simplifications are typically made.  First, it is assumed that the polypeptide chain can be represented using only the $C_{\alpha}$ atoms.  Second, the connected $C_{\alpha}$ atoms are confined to the vertices of a suitable lattice.  A large number of studies have been done using square or cubic lattices\cite{DillProtSci95}.  { }Fig.\ \ref{fig:1}{\bf A} provides an illustration of a model of a polypeptide chain in a cubic lattice. { }To satisfy the excluded volume condition, only one bead is allowed
to occupy a lattice site. The energy of a conformation, specified by \(\left\{\mathbf{r}_i\right\}_{i=1}^N\), is 

\begin{equation}
E\left[\left\{\mathbf{r}_i\right\}\right]=\sum _{i<j+3} \Delta \left[\left|\mathbf{r}_i-\mathbf{r}_j\right|-a\right]B_{ij},
\label{eqn:1}
\end{equation}

\noindent where $N$ is the number of beads in the chain, $a$ is the lattice spacing, and \(B_{ij}\) is the value of the contact interaction between beads
i and j. { }Typically, the lattice spacing is chosen to correspond to the distance between $C_{\alpha} $ atoms along the protein backbone ( $\simeq
$ 3.8 {\AA} ). { }Several different forms for the elements of the contact matrix, \(B_{ij}\), have been used in many studies.
Note that the discrete form of the energy makes it suitable only for Monte-Carlo simulation or exact enumeration provided that N is small (i.e., N $\approx$ 25 for square lattice and N $\approx$ 20 for cubic lattice).  

Despite the drastic simplifications, great insight into global folding mechanisms were obtained using lattice models.  Their success in providing a caricature of the folding process can be attributed to their ability to capture certain global protein-like properties.  The two most salient features of native protein structures are that they are compact and that protein interiors are largely made up of hydrophobic residues, while hydrophilic residues are better accommodated along a protein's surface.  Foldable lattice sequences capture these features, and allow us to exhaustively simulate the thermodynamics and folding kinetics even when $N$ is relatively large.  Here, we give a few examples from simulations of lattice models that provided a conceptual understanding of the structure of sequence space and folding mechanisms, as well as specific predictions for the role of disulfide bonds in the folding process.

\subsubsection*{Thinning of Sequence Space} 
One of the interesting uses of lattice models was to provide a plausible physical basis for the thinning of sequence space, leading to foldable sequences.
The Hydrophic-Polar (HP) model \cite{LauMacro89, ChanJCP89, DillProtSci95} (two-letter code for amino acids), and exact enumeration of all possible conformations of the HP model, with $N \leq 25$, allowed Camacho and Thirumalai \cite{CamachoPRL93} to 
confirm the well known results that the number of self-avoiding conformations and maximally compact structures grow exponentially with N. Because a protein's folded structure is
unique, it follows that specific native interactions 
among the large number of compact structures are required to mimic protein-like behavior.  Surprisingly, it was found that the number of minimum energy compact structures (MECS) increased very slowly with N.  It was conjectured that $C_{MECS}(N) \sim \ln(N)$.  The generality of these results were subsequently confirmed using the three dimensional random bond (RB) model \cite{KlimovProteins96}, in which elements of the interaction matrix \(B_{\text{ij}}\) were distributed according
to a Gaussian with mean 0 or -0.1 (55$\%$ of residues are hydrophobic) and unit variance.{ }Thus, imposing minimal restrictions on the protein structures (compactness and low energies) naturally leads to a sparse structure space.  The clear implication of $C_{MECS}(N) \sim \ln(N)$ is that many sequences likely map onto the same fold.  
In an important article, Li et al.\ \cite{LiScience96} explicitly showed that this is indeed the case by considering 27-mer HP
models on a 3D cubic lattice. { }They found using simulations, that certain putative basins of attraction in the fold space served as attractors for a much larger number of sequences than
others; such structures were considered more designable than others. { }Lindgard and Bohr \cite{LindgardPRL96}  further substantiated these ideas by demonstrating that 
only very few compact folds are compatible with protein characteristics.  All of these studies confirmed that
the density of the structure space is sparse, and that each natural fold can be designed by many sequences. We parenthetically note that recent single molecule 
experiments, that use force-quench to initiate folding, have clearly revealed the role of MECS in directing protein folding from unfolded conformations (J. M. Fernandez, private communication).

\subsubsection*{Foldability and Folding Mechanisms}
Although many sequences map onto the same structure, not all such sequences are viable as proteins. { }This is because of the dual requirements
of thermodynamic stability and kinetic competence. { }Not only must a native protein be compact and be of low energy, but (under folding conditions) 
it must be able to adopt such a state on a biologically relevant time scale.  Lattice models have also proven useful in understanding the kinetics of protein folding.  
Using the HP model\cite{CamachoPNAS93} and subsequently the RB model\cite{Klimov02JMB}, it was shown that the parameter 
\begin{equation}
\label{eqn:sigmaCT}
\sigma_{CT} = (T_{\theta} - T_{F})/T_{\theta}
\end{equation}
\noindent governs the foldability of proteins.  They considered several 27 bead sequences and computed the mean first passage time, $\langle$ \(\tau _f\) $\rangle$, to the native conformation. { }Small changes in the value of \(\sigma_{CT}\) resulted in dramatic (a few orders of magnitude) increases in \(\tau _f\).  Thus, the dual requirements of stability and kinetic accessibility
of the folded state are best satisfied by those sequences that have relatively small values of \(\sigma _{CT}\).  Similarly, lattice simulations have also shown \cite{SocciJCP96} that foldable sequences have $T_{F}/T_{g} \approx 1.6$.  Here, $T_g$ is a kinetic glass transition temperature that is signaled by a large increase in the optimal folding time.  
These studies also provide numerical evidence for the relationship between $T_{g}$, $T_{F}$, and $T_{\theta}$.

Lattice models have even been used to qualitatively understand the mechanisms for the folding of proteins, and in particular the kinetic partitioning
mechanism (KPM) \cite{ThirumalaiCOSB99}. { }  According to KPM, a fraction of proteins $\Phi $ reach the NBA rapidly without being trapped in other competing
basins of attraction (CBA). { }Only on longer time scales do fluctuations allow CBA trapped molecules to reach the NBA. { }$\Phi $ is sequence dependent
and explicitly determined by \(\sigma _{CT}\). { }Thus, in general we can write the time dependence of the fraction of molecules that have not folded
at time $t$, \(P_u\)(t), as,
\begin{equation}
\label{eqn:2}
P_u(t)=\Phi \exp{\left(-t/\tau_F\right)}+\sum _k a_k\exp{\left(-t/\tau_k\right)},
\end{equation}
\noindent where \(\tau _{\text{F}}\) is the time constant associated with the fast-folders, \(\tau _k\) is the escape time from the CBA labeled k, and
\(a_k\) is the fraction of molecules initially trapped in the $k$-th CBA. 
The simulations using off-lattice and lattice models showed that some trajectories reach the native
state directly from random coil conformations, while others get trapped and require much longer times to reach the native state.  The validity of KPM has been firmly established for the folding of both proteins and RNA\cite{ThirumalaiACR96}.

\subsubsection*{Disulfide Bonds in Folding}
The refolding pathways of bovine pancreatic trypsin inhibitor (BPTI) were intensely scrutinized, because BPTI's native state can be characterized
by three disulfide bonds between Cys residues.  The native conformation of the 56 residue BPTI contains three
disulfide bonds between residues 5,14,30,38,51, and 55, and is denoted by [30-51;5-55;14-38]. Experiments show that, under oxidizing conditions, the native state is formed reversibly from among the 75 possible intermediates consisting of single, double, and triple disulfide bonds.  
The refolding pathways are characterized by the accumulation of the various intermediates.  Early experiments showed that of the 75 possible intermediates
only eight could be detected on the experimental time scale \cite{Creighton92Science,Creighton84JMB,Creighton77JMB}.  Most importantly, it was shown that three non-native states, the intermediates 
with disulfide bonds not present in the native state - are well populated.  In particular, the non-native species [30-51;5-14] and [30-51;5-38] were involved
in the productive pathway; this is, folding proceeds through either of these two kinetically equivalent intermediates.  The key role of non-native intermediates
in driving BPTI folding was challenged by Weissman and Kim (WK) \cite{Weissman91Science,Weissman92PNAS}, who used a rapid acid quenching method to disrupt the folding process, and determined 
the nature of populated intermediates.  Based on these studies, it was argued that, in the productive pathway, only native intermediates play a significant role.  Non-native intermediates may only be involved as required
by disulfide chemistry in the last stages of the folding of BPTI; that is, they play a role in the formation of the precursor [30-51;5-55] from [30-51;14-38] (denoted by 
$N_{sh}^{sh}$ and $N^\prime$, respectively).

To resolve the apparent controversy between the distinct proposals for BPTI folding\cite{CamachoPROTEINS95}, we introduced a theory based on
the proximity rule and simulations based on lattice models\cite{CamachoPNAS95}.  The basic concept of the proximity rule is that local events, governed
largely by entropic considerations, determine the initial folding events.  Because the conformations of the intermediates that determine the folding pathway
are specified in terms of S-S bonds in BPTI, they can be used as a surrogate reaction coordinate.  There are two ingredients in the proximity rule:
(1) Under oxidizing conditions the probability of the S-S formation is viewed as forming loops of appropriate length.  The probability of forming a loop of length
$l$ is\cite{ThirumalaiJPCB99} 
\begin{equation}
\label{eqn:3}
P(l) \approx [1 - e^{(-l/l_p)}]/l^{\theta _3}
\end{equation}
\noindent where $\theta _3 \approx 2.2$, and $l_p$ is the persistence length of the protein.
(2)  The second component of the proximity rule is related to the kinetics of native state formation.  Folding follows a three-stage kinetics \cite{CamachoPNAS93}.  (i)  There is a rapid
collapse of the chain to a set of compact conformations driven by a competition between hydrophobic forces and loop entropy.  In BPTI, this is characterized
by the formation of loops between Cys residues, so that the S-S bonds form.  At the end of this stage, the most stable single disulfide species accumulate.  (ii)  In the second
stage, intermediates with single disulfide bonds rearrange to form native two-disulfide species.  (iii)   The rate-determining step involves the transition from the stable
two-disulfide species to the native conformation.  In this sequential progression bifurcations in the folding pathways are possible resulting in the parallel pathways to the native
state \cite{CamachoPNAS95}.

The proximity rule and experimentally determined times for rearrangement of single S-S intermediates to double S-S species were use to map out the refolding
pathways\cite{CamachoPNAS95}.  The theory showed that, on long time scales, only native-like species are populated, which is in accord with the WK experiments.  In the process
of formation of $N_{sh}^{sh}$ and $N^\prime$, it is likely that non-native intermediates form transiently.  The key prediction of our theory was 
that the native single disulfide intermediate [14-38] forms rapidly in the folding process.  However, the instability of the intermediate [14-38] results in a decrease in its concentration
where as those of the metastable native species [30-51] and [5-55] increase.  The theoretical prediction was subsequently confirmed by Dadlez and Kim\cite{DadlezBiochem96} who showed using oxidized glutathione that [14-38]
is the first intermediate to form.  The confirmation of this key theoretical prediction validates the proximity rule, and the general principle that native interactions should dominate
the folding process even if non-native species accumulate transiently early in the folding process.
 
\begin{figure}[ht]
\includegraphics[width=7.00in]{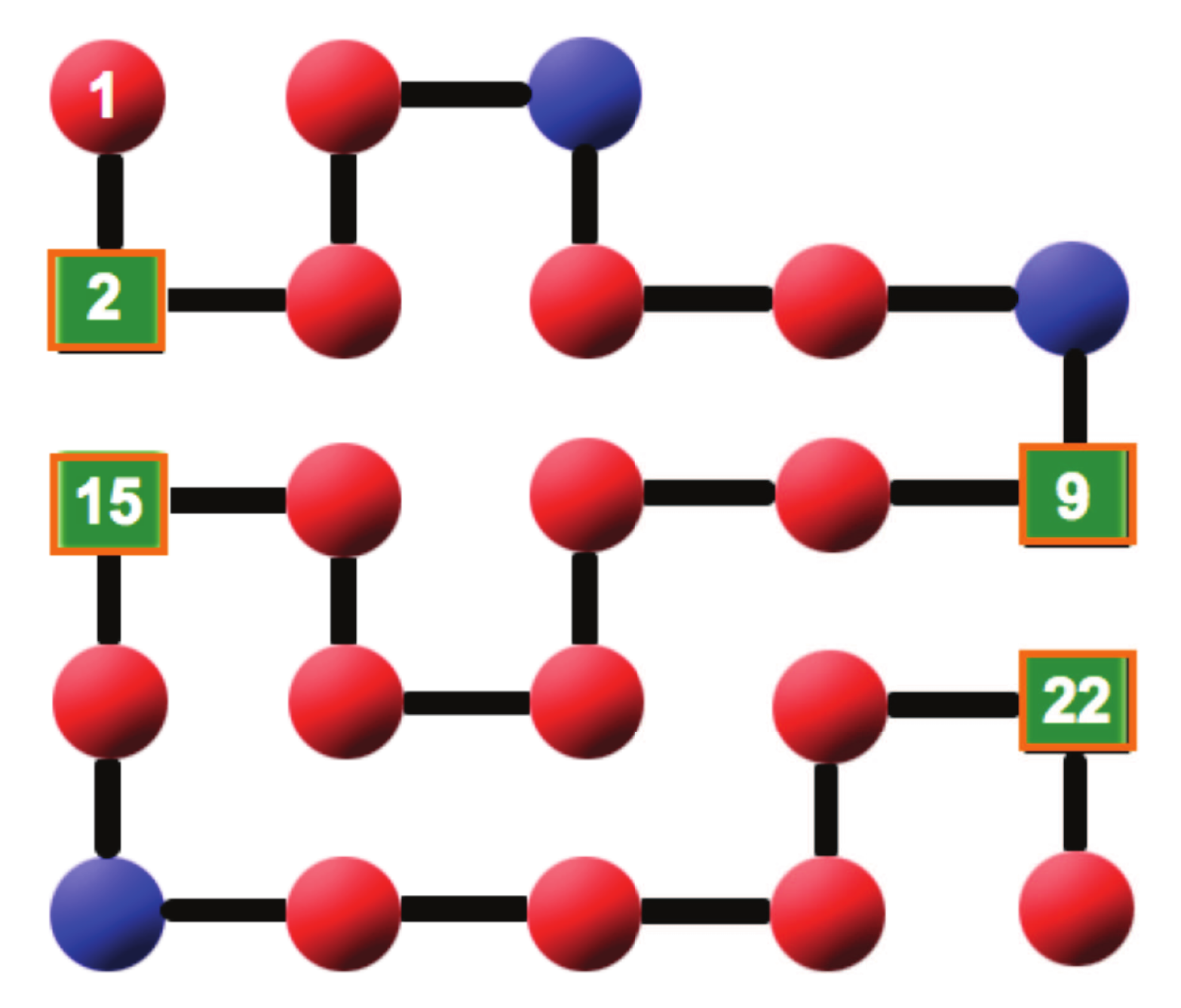}
\caption{}
\label{fig:2}
\end{figure}

To further highlight the relevance of non-native intermediates in the folding of proteins we used simple lattice models with disulfide bonds \cite{CamachoPROTEINS95}.  A two-dimensional lattice
sequence consisting of hydrophobic (H), polar (P), and Cys (C) residues was simulated to incorporate the role of S-S bonds.  If two beads are near neighbors on the lattice,
they can form a S-S bond with associated energy gain of $-\epsilon _s$ with $\epsilon _s > 0$.  This model is a variant of the HP model in which $\epsilon_{s} / \epsilon_{h}$ ($\epsilon_{h}$ is the interaction strength between the hydrophobic residues) controls the refolding pathway.  Because of the simplicity of the model, it can be simulated in great detail
to provide insights into the role of non-native intermediates in S-S driven folding.  We considered a sequence of $M = 23$ monomers, of which four represent C sites.
The native conformation is specified as [2-15;9-22].  The model sequence has six possible single and two disulfide intermediates including the native state (Fig.\ \ref{fig:2}).  There are
three native intermediates and two non-native intermediates.

\begin{figure}[ht]
\includegraphics[width=7.00in]{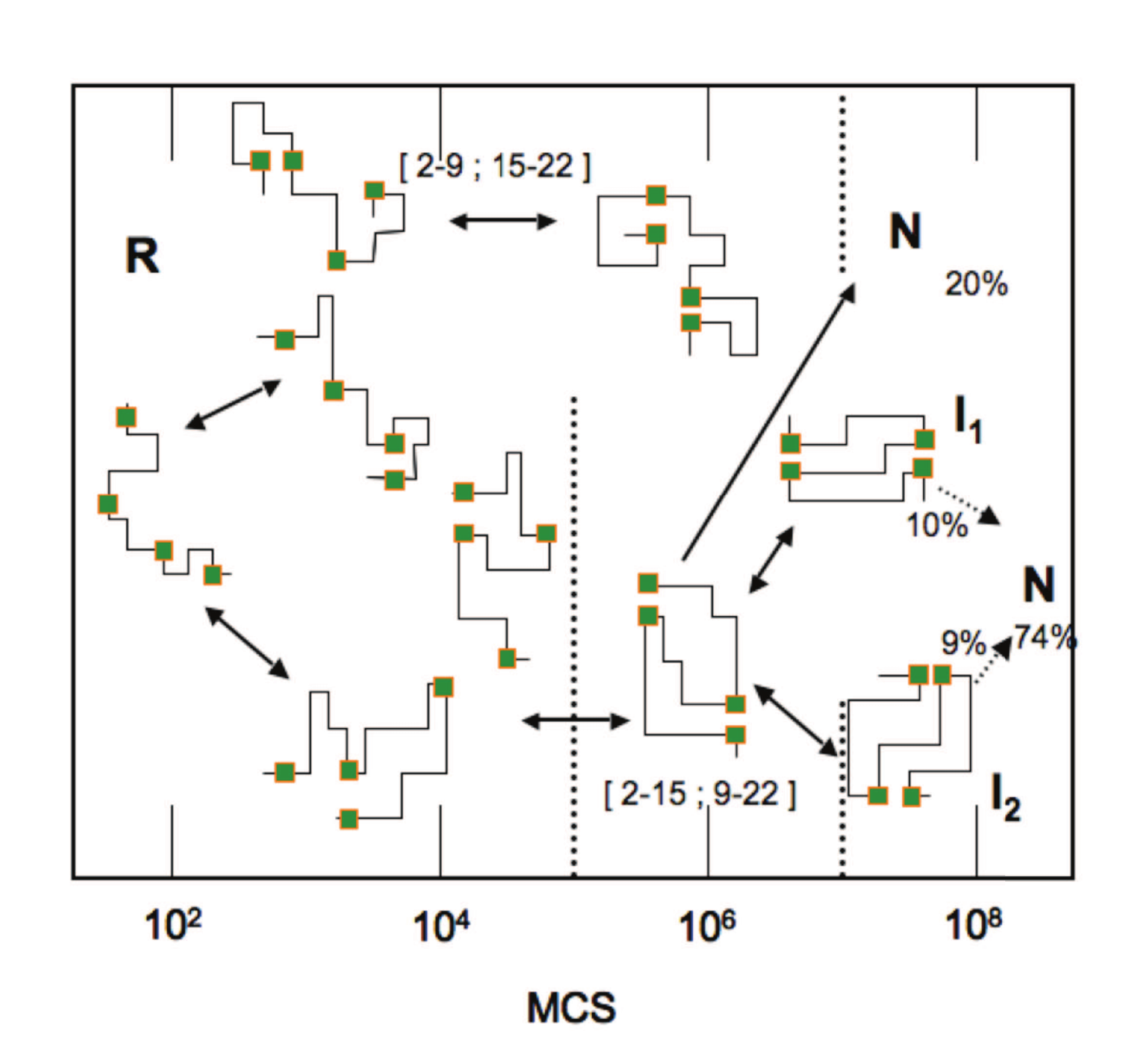}
\caption{}
\label{fig:3}
\end{figure}

The folding pathways in Fig.\ \ref{fig:3} are characterized by the time dependent concentration of the six species.  Even in this simple model, 
the routes leading to the native state (N) shows evidence for
multiple routes.  Clearly, there are pathways that reach N exclusively via native-like intermediates.  In other routes, non-native intermediates are populated early in the 
folding process.  At the shortest times, (measured in units of Monte Carlo Steps) $t < 10^{-5} \tau_f$ ($\tau_f$ is the folding time) single disulfide bonds accumulate.  
The probability of their formation can be predicted using $P(l)$ (Eq.\ (\ref{eqn:3})).  When $t \approx 10^{-1} \tau_f$ the non-native single disulfide intermediates rearrange to form the more
stable native [9-22] and [2-15] species.  Their stabilities are determined by entropy loss due to the favorable enthalpic gain arising from hydrophobic collapse.  The single
disulfide species act as nucleation sites for further consolidation of the native state.

In the second stage of the assembly a non-native two-disulfide intermediate [2-9;15-22] forms transiently.  Because this intermediate is unstable it quickly
rearranges to the more stable native {\bf N} state.  Interestingly when $t \approx 0.01 \tau _F$ there are two native-like intermediates, in which the disulfide bonds
are in place but some other parts are not fully structured.  This is the analogue of the $N_{sh}^{sh}$ state in BPTI which only needs the nearly solvent-exposed [14-38]
bond to form to reach N.  In the final stage of folding, structural fluctuations that transiently break the native S-S bonds enable the transition to {\bf N}.  The transition involves
transient formation of the non-native intermediate [2-9;15-22].  The two native-like intermediates $I_1$ and $I_2$ (Fig.\ \ref{fig:3}) rearrange almost exclusively through the native intermediate.

Several important conclusions for BPTI folding emerged from this study.
(1) Non-native species form early in the folding process when the ordering is determined by entropic considerations.  The current experiments
on BPTI are far too slow to detect these intermediates.  On the time scale of collapse, stable native-like single disulfide species form.  This study also justifies the
use of models that emphasize the role of native-interactions in driving the folding process.  The structure based models, that discourage non-native
contact formation, probably only quantitatively influences the description of the earliest events in the folding process.  In order to obtain an accurate description of
such processes a detailed characterization of the denatured state ensemble, in which non-native interactions may play a role, is required.
(2)  As the folding reaction progresses, native-like intermediates form so that the productive pathways largely contain native-like intermediates.
(3)  The rate-determining step involves an activated transition from native-like species, via a high free-energy non-native transition state
to {\bf N}.  The transition involves rearrangement of the structure that does not involve the S-S bonds.  We concluded that, although the folding
pathways of BPTI can be described in terms of disulfide intermediates, a complete description requires accounting for hydrophobic and charge effects as
well.  The profound effect of point mutations in altering the folding rates and the pathways of BPTI folding\cite{ZhangBiochem93} suggests that there are strong couplings
between S-S bond formation and other forces that drive the native structure formation.  These findings are in accord with predictions
using lattice models\cite{CamachoPROTEINS95,ThirumalaiARBBS01}.

\subsection*{Folding Using Off-Lattice Models}
Since the earliest efforts of Flory to understand conformational
transitions in peptides, there has been considerable effort
to develop off-lattice models for proteins.  The pioneering work of
Levitt and Warshel\cite{LevittNature75} showed that some aspects of refolding
of BPTI can be captured using a simplified representation of polypeptide
chains. This work, which in retrospect should be viewed as the first
attempt to simulate globular folding using CG models, has laid the foundation for
devising various off-lattice models. Spurred in part by theoretical
arguments
(see Introduction), Honeycutt and Thirumalai (HT)\cite{HoneycuttBP92} developed a
$C_{\alpha}$-representation of polypeptides for which rigorous
simulations of
thermodynamics and kinetics could be carried out.  The HT model and
variations have formed the basis of numerous studies that
have profitably
been used to gain insights into a number of increasingly complex
problems.  By using a few examples, we illustrate the power of these
models and the need to
develop appropriate variations depending on the problem at hand.

\subsubsection*{$C_{\alpha}$ Models} The original HT model \cite{HoneycuttBP92}, which used 
a three letter representation (hydrophobic, polar, and neutral) of polypeptides, was used to probe the
energy landscapes of $\beta$-barrel structures. The typical energy
function used in the simulations
of $C_{\alpha}$-models is given by
\begin{equation}
\label{eqn:4}
V_{\alpha} = \sum_{i=1}^{N _\theta}V(\theta _i) + \sum_{i=1}^{N _\phi}V(\phi _i) + \sum_{j > i+3}V_{ij},
\end{equation}
\noindent where $V(\theta _i) = \frac{k _\theta}{2}(\theta _i - \theta_0)^2$ and $V(\phi _i) = A(1 + \cos\phi _i) + B(1 + \cos(3\phi _i))$.
Thus, bond angles are harmonically constrained about equilibrium values
of $\theta _0 = 105^{\circ}$ and the torsion potential contains three minima ( a global 
minimum corresponding to the trans-state and two slightly higher gauche minima ).

Two hydrophobic beads interacted via the following attractive potential:
\begin{equation}
\label{eqn:5}
V_{ij} = 4\varepsilon _h\left[\left(\frac{\sigma}{r_{ij}}\right)^{12}-\left(\frac{\sigma}{r_{ij}}\right)^6\right]. 
\end{equation}  
\noindent Neutral beads interacted with neutral, hydrophilic, and hydrophobic beads via the short-ranged repulsive potential:
\begin{equation}
\label{eqn:6}
4\varepsilon _h\left(\frac{\sigma}{r_{ij}}\right)^{12}.
\end{equation}
Hydrophilic beads interacted with hydrophilic and hydrophobic beads via a longer-ranged repulsive potential:
\begin{equation}
\label{eqn:7}
4\varepsilon _l\left[\left(\frac{\sigma}{r_{ij}}\right)^{12}+\left(\frac{\sigma}{r_{ij}}\right)^6\right].
\end{equation}

Using this model, HT computed the spectrum of low energy conformations
that showed that the native state is separated by
an energy gap from native-like structures.  However, the interplay between the entropy of the native-like
conformations and
the energy gap\cite{KlimovJCP98}, that can be succinctly expressed in terms of the
characteristic temperatures, determines foldability\cite{ShakhnovichBiophysChem1989}.

\begin{figure}[ht]
\includegraphics[width=7.00in]{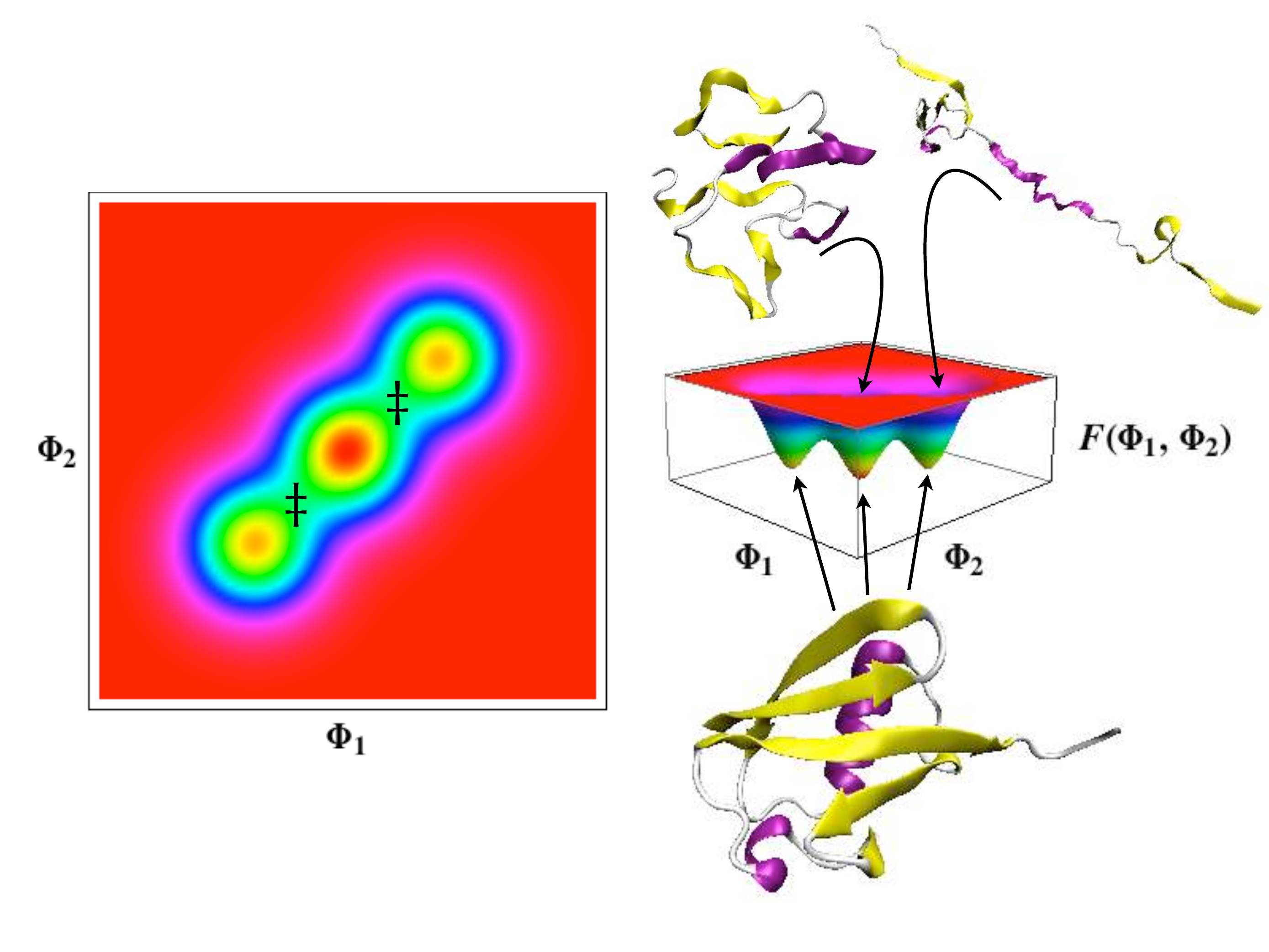}
\caption{}
\label{fig:4}
\end{figure}

An important result in the HT study is
that sequences that are topologically and energetically frustrated can
be trapped in native-like conformations for prolonged periods of time.
Such conformations, which are functionally competent and kinetically
accessible would render them metastable (Fig.\ \ref{fig:4}).  While many foldable
sequences do not fall into this category,  the metastability
hypothesis is important in the context of aggregation-prone proteins.
For example, it has been suggested that the normal cellular form of
the mammalian prion protein, PrP$^{C}$ may
well be metastable because regions of the C-terminal ordered
structure are frustrated \cite{Dima04PNAS}.

The energy landscape of the HT model is rugged. Indeed, refolding in
such a landscape occurs by the KPM\cite{GuoBP95} (see Eq.\ (\ref{eqn:2})). While such a model
accurately describes the folding of lysozyme\cite{KiefhaberPNAS1995}, there are a number of examples in
which folding occurs by two-state kinetics. Because the folding
landscape of such proteins
is relatively smooth, it was realized that upon elimination of
non-native interactions the folding efficiency could be enhanced. With
this observation and the notion that
native topology drives folding Clementi et. al \cite{ClementiJMB00,KaranicolasProtSci02}, devised
structure-based Go models. In this class of models, the energy function
is a variation of the one given in Eq.\ (\ref{eqn:4}) except
that interactions that are not present in the native state are
repulsive.  The resulting $C_{\alpha}$-Go model has been used with success
in probing the refolding of a large number of experimentally
well-characterized proteins (e.g., CI2\cite{ClementiJMB00}, SH3 domain\cite{ClementiJMB00}, and Interlukin\cite{Gosavi06JMB}).  These
studies clearly show that simple models, with physically motivated
approximations, provide valuable insights into protein folding
kinetics.

\subsubsection*{$C_{\alpha}$-SCM}  It is well known that, although proteins can
tolerate large volume mutations in their core without being fully
destabilized, their interior
is densely packed. Indeed, a detailed analysis of the shapes of folded
structures shows that single domain proteins are highly spherical\cite{Dima04JPCB}. In
order to capture the packing
of the largely hydrophobic core, it is important to go beyond the
simple $C_{\alpha}$ models.  In addition, studies using lattice models
with side chains showed that
the extent of cooperativity is better captured if the interior is densely
packed\cite{KlimovFoldDes98}. To provide a more realistic representation, Klimov and
Thirumalai \cite{KlimovPNAS00} represented a polypeptide chain using two interaction sites
per amino acid residue (except Gly). One of the sites is the
$C_{\alpha}$ atom and the other represents the side chain.  The sizes
of the side chains were taken to be proportional to their van der
Waals radii. The resulting $C_{\alpha}$-SCM was first applied to study
the formation of a $\beta$-hairpin.  To date this is the only study
whose results quantitatively agree with thermodynamic measurements\cite{MunozNat97} 
and measurements of its folding kinetics. More importantly, they also showed that
the transition to the ordered structure occurs over a very broad
temperature range due to finite-size (16 residues) of the system.  \emph{In
silico} mutational studies also showed that the mechanism of hairpin
formation, that involves an interplay of collapse and turn formation,
depends on the loop stiffness. This result, which was further
developed using $\Phi$-value analysis, was used to propose that the
stiffness of the distal loop in the SH3 domain leads to a polarized
transition state in its folding\cite{Klimov02JMB}.

There are a variety of novel applications using the $C_{\alpha}$-SCM.
Most noteworthy is the use of these models to probe the effects of
molecular crowding on the stability and folding kinetics of WW domain,
an all $\beta$-sheet protein.  By modeling the crowding particles as
spheres Cheung and Thirumalai \cite{CheungJPCB07} showed that crowding enhances the
stability of the protein relative to the bulk. The folding rates also
increase non-monotonically as the volume fraction is increased. These
results were explained theoretically by approximately mimicking
crowding effects by confinement.  More recently,
Cheung and coworkers have extended these treatments to larger
proteins\cite{StaggPNAS07, Homouz08PNAS}. In collaboration with experimentalists, they have have shown
that the ideas developed in the context of the small WW domain also
apply to larger systems. These impressive simulations further
illustrate the use of $C_{\alpha}$-SCM in the study of problems that
are realistic models for folding under cellular conditions.

\subsubsection*{Self-Organized Polymer (SOP) Model for Single Molecule Force Spectroscopy}
The remarkable progress in using C$_\alpha$ models and C$_\alpha$-SCM models has, in general, been restricted to relatively small proteins ($N \sim 100$ residues).  
For $N$ much larger than about 100 converged simulations become difficult to carry out, even for minimal models.  However, many of the problems of current interest,
such as protein-protein interactions, links between allosteric transitions and protein function, and movements in molecular machines often involve
thousands of residues.  In order to tackle a subset of these problems, we have devised a class of models that is even simpler to simulate than the well 
known C$_\alpha$ and C$_\alpha$-SCM models. The resulting model has to be realistic enough to take into account the interactions
that stabilize the native fold, yet be simple enough that within finite computational time one can trace the transition dynamics of large
molecules. The self-organized polymer (SOP) model \cite{Chen07JMB,Mickler07PNAS,Hyeon06PNAS,Hyeon07PNAS,Hyeon07PNAS2,HyeonPNAS05,Hyeon06Structure,HyeonBJ07}, 
a prototype for a new class of versatile coarse-grained structure-based models, is well suited to understanding 
dynamics at the spatial resolution that single-molecule force spectroscopy of large proteins provides.  

We have recently introduced the
SOP model to study the response of proteins and RNA to mechanical force \cite{Mickler07PNAS,Hyeon06Structure,HyeonBJ07}.  The reason for using the SOP model 
in force spectroscopy applications is the following: (i) Forced-unfolding and force-quench refolding lead to large conformational changes
on the order $\sim $ (10-100) nm. Currently, single molecule experiments (laser optical tweezers or atomic force microscopy) cannot resolve structural changes
below 1 nm \cite{Bustamante2,Bustamante03Science,FernandezSCI04,FernandezNature99}. As a result, details of the rupture of hydrogen bonds or local contacts between specific residues cannot be discerned from FEC's or
the dynamics of the end-to-end distance ($R$) alone. Because only large changes in $R$ are monitored, it is not crucial to model minor perturbative details 
due to  local interactions such as bond-angle and various dihedral angle potentials. As shown in the literature on normal-mode models\cite{BaharCOSB05}, the inclusion
of small details only affects the higher frequency modes, and the global dynamics are mainly determined by the low frequency normal modes\cite{BaharPRL97,BaharCOSB05,Zheng06PNAS}.
Such modes, that are linked to function, are robust\cite{Zheng06PNAS} as long as the topological constraints are not altered. 
(ii) In the context of mechanical unfolding as well as the folding of proteins, many of the
details of the unfolding and folding pathways can be accurately computed by taking into account only the interactions that stabilize the native fold \cite{KlimovPNAS00}.
Previous studies also suggested that it is crucial to take into account chain connectivity and attractive interactions that faithfully reproduce
the contact map of a fold.  The basic idea of the SOP model is to use the simplest possible Hamiltonian to simulate the low-resolution global dynamics 
for proteins of arbitrary size.  The energy function for proteins in the SOP representation of polypeptide chains is 
\begin{equation}
\label{eqn:8}
\begin{split}
V_{\text{SOP}}&=V_{\text{FENE}}+V_{\text{NON}}\\
 &=-\sum _{i=1}^{N-1} \frac{k}{2}R_0^2\log\left[1-\frac{\left(r_{i,i+1}-r_{i,i+1}^0\right)^2}{R_0^2}\right] + \sum _{i=1}^{N-3} \sum _{j=i+3}^N
\varepsilon _h \left[\left(\frac{r_{i,j}^0}{r_{i,j}}\right)^{12}-2 \left(\frac{r_{i,j}^0}{r_{i,j}}\right)^6\right]\Delta _{\text{ij}}\\
& \quad +\sum _{i=1}^{N-3}
\sum _{j=i+3}^N \varepsilon _l\left(\frac{\sigma }{r_{i,j}}\right)^6\left(1-\Delta _{\text{ij}}\right)
+\sum_{i=1}^{N-2} \varepsilon _l\left(\frac{\sigma }{r_{i,i+2}}\right)^6
\end{split}
\end{equation}
The first term in Eq.\ (\ref{eqn:8}) is the finite extensible nonlinear elastic (FENE) potential for chain connectivity with 
parameters, $k = 20 $ kcal/(mol \AA$^2$), $R_0 = 0.2$ nm, $r_{i,i+1}$ is the distance between neighboring beads at i and i+1, and $r_{i,i+1}^0$ is the distance in
the native structure. The use of the FENE potential is more advantageous than the standard harmonic potential, especially for forced-stretching, because
the fluctuations of $r_{i,i+1}$ are strictly restricted around $r_{i,i+1}^0$with variations of $\pm R_0$ to produce worm-like chain behavior. The Lennard-Jones
potential is used to account for interactions that stabilize the native topology. A native contact is defined for bead pairs $i$ and $j$ such that $|i-j|> 2$ 
and whose distance is less than 8 \AA{ }in the native state. We use $\varepsilon _h = 1-2$ kcal/mol for native pairs, and $\varepsilon _l = 1$ kcal/mol for nonnative pairs.
In the current version, we have neglected nonnative attractions.  This should not qualitatively affect the results, because under tension such interactions
are greatly destabilized. To ensure noncrossing of the chain, i,i+2 pairs interacted repulsively with $\sigma =3.8$ {\AA}. There are five parameters
in the SOP force field. In principle, the ratio of $\varepsilon _h/\varepsilon _l$ and $R_c$ can be adjusted to obtain realistic values of critical
forces. For simplicity, we choose a uniform value of $\varepsilon _h$ for all protein constructs.  $\varepsilon _h$ can be made sequence-dependent
and ion-implicit if one wants to improve the simulation results. 
 
The time spent in calculating the Lennard-Jones forces scales as $\sim O(N^2)$. Drastic savings in computational time can be achieved by
truncating forces due to the Lennard-Jones potential for interacting pairs with $r_{ij} > 3r_{ij}^0$ or $3\sigma$ to zero. We refer to the model as the `self-organized polymer'
(SOP) model because it only uses the polymeric nature of the biomolecules and the crucial topological constraints that arise from the specific fold.
For probing forced-unfolding of proteins (or RNA), it is sufficient to only include attractive interactions between contacts that stabilize the native
state. We believe none of the results will change qualitatively if this restriction is relaxed, i.e., if nonnative interactions are also taken into
account.

\subsection*{Forced-Unfolding and Force-Quench Refolding of GFP}
Recently, single molecule force experiments using AFM have been exploited to unravel GFP from its native structure.  The measured force-extension
curves (FEC's) were used to construct its partial energy landscape \cite{DietzPNAS04}. Two unfolding intermediates were identified; 
the first intermediate (GFP$\Delta \alpha $) results from the disruption of H1 (Figure \ref{fig:5}), and the second, GFP$\Delta \alpha \Delta \beta $, 
was conjectured to be either unraveling of $\beta
$1 from the N-terminus or $\beta $11 from the C-terminus. Precise assignment of the structural characteristics of the intermediate is difficult not
only because of the complex topology of GFP but also because, unlike in RNA, secondary structures in proteins are typically unstable in the absence
of tertiary interactions.  Thus, it is impossible to obtain the unfolding pathways from the FEC alone.

\subsubsection*{Mechanical Unfolding of GFP} 
\begin{figure}[ht]
\includegraphics[width=7.00in]{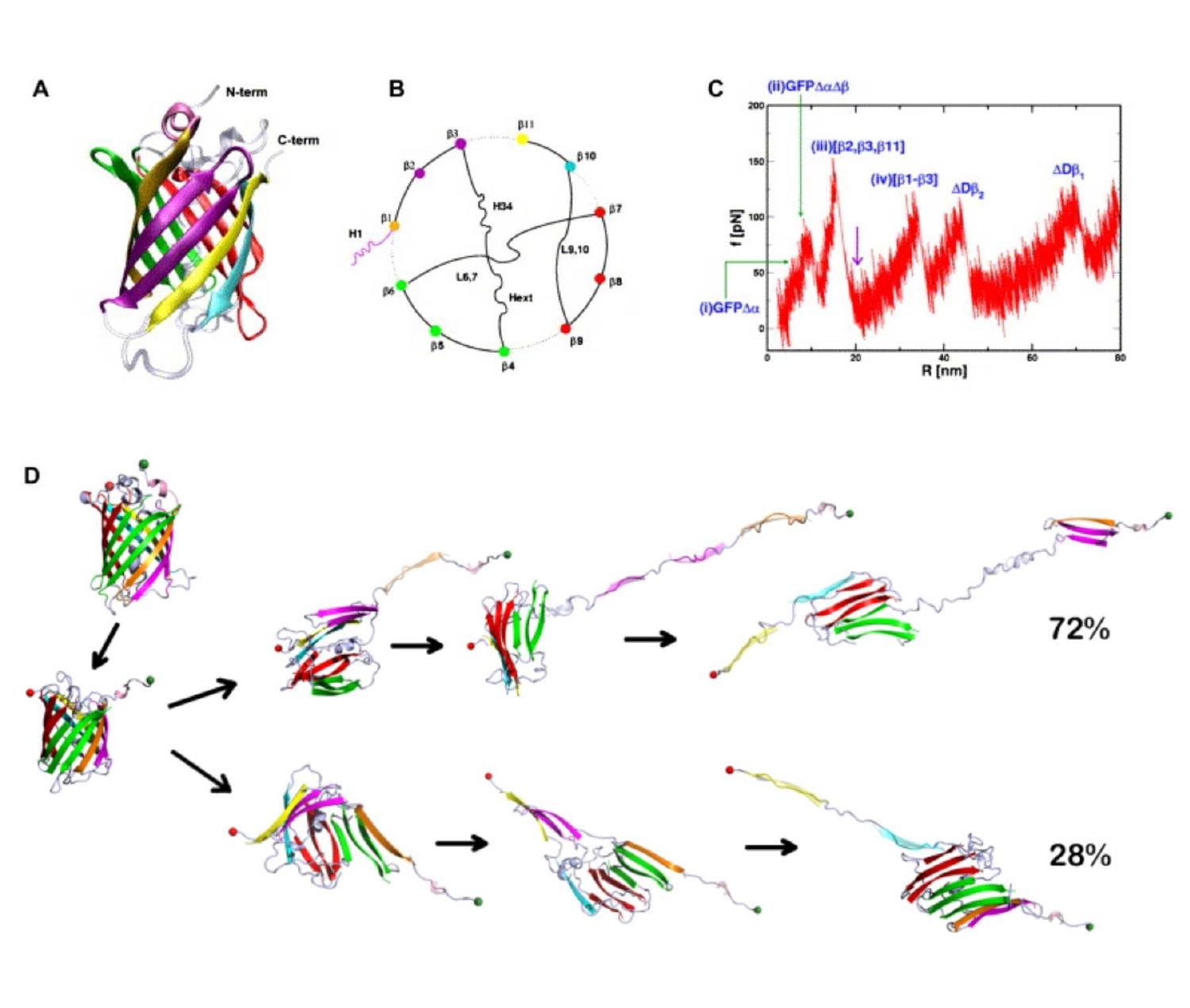}
\caption{}
\label{fig:5}
\end{figure}

The native state of GFP (PDB file 1gfl in Figure \ref{fig:5}{\bf A}) consists of 11 $\beta $-strands, three helices,
and two relatively long loops.  A two-dimensional connectivity map of the $\beta $-strands shows that $\beta $4, $\beta $5, $\beta $6 and $\beta
$7, $\beta $8, $\beta $9 are essentially disjointed from the rest of the structure (Fig.\ \ref{fig:5}{\bf B}).  From the structure alone, we expect that the strands
in the substructures (D$\beta _1$ { }[$\beta $4,$\beta $5,$\beta $6]) and (D$\beta _2$ [$\beta $7,$\beta $8,$\beta $9]) would unravel almost synchronously.  
We probed the structural changes that accompany the forced-unfolding of GFP using FEC's and the dynamics of rupture of contacts at $v = 2.5 \mu$m/sec
( $2.5 v_\text{AFM}$ ), where $v$( $v_\text{AFM}$ ) is the pulling speed ( pulling speed used in AFM experiments ). 
The unfolding FEC's in a majority of molecules have several peaks (Fig.\ \ref{fig:5}{\bf C}) that represent unfolding of the specific secondary structural
elements (SSE's). By using simulations to monitor contact (residue-residue) rupture, the structures that unravel can be unambiguously assigned to
the FEC peaks. Unfolding begins with the rupture of H1 (leading to the intermediate GFP$\Delta \alpha $), which results in the extension by about $\Delta
$z $\approx$ 3.2 nm (Fig.\ \ref{fig:5}{\bf C}).  The force required to disrupt H1 is about 50 pN (Fig.\ \ref{fig:5}{\bf C}) which compares well with the experimental estimate of 
35 pN \cite{DietzPNAS04}.  In the second intermediate, GFP$\Delta \alpha \Delta \beta $, $\beta $1 unfolds \cite{DietzPNAS04}. The value
of the force required to unfold $\beta $1 is about 100 pN (Fig.\ \ref{fig:5}{\bf C}), which is also roughly in agreement with experiment \cite{DietzPNAS04}. After
the initial events, the unfolding process is complex. For example, ruptured interactions between strands $\beta $2 and $\beta $3 transiently reform
(Fig.\ \ref{fig:5}{\bf D}). The last two rips represent unraveling of D$\beta _1$ and D$\beta _2$ in which the strands in D$\beta _1$ and D$\beta _2$ unwind nearly
simultaneously.

Besides the dominant pathway (72$\%$) described above (Fig.\ \ref{fig:5}{\bf D} top), a parallel unfolding route is navigated by some of the trajectories \cite{Mickler07PNAS}.
In the alternative pathways (28$\%$) (Fig.\ \ref{fig:5}{\bf D} bottom) the C-terminal strand $\beta $11 unfolds after the formation of GFP$\Delta \alpha $. In
both the dominant and the subdominant routes, multiple intermediates are observed in simulations. To assess if the intermediates in the dominant pathway
are too unstable to be detected experimentally, we calculated the accessible surface area of the sub-structures using the PDB coordinates for
GFP. The structures of the intermediates are assumed to be the same upon rupture of the SSEs, and hence our estimate of surface area is a lower bound.
The percentage of exposed hydrophobic residues in the intermediate [$\beta $2,$\beta $3,$\beta $11] is 25$\%$ compared to 17.4$\%$ for the native
fold whereas in excess of 60$\%$ of the hydrophobic residues in $\Delta $D$\beta _2$ are solvent accessible. We conclude that the intermediate [$\beta
$2,$\beta $3,$\beta $11] in which H1, $\beta $1-$\beta $3, and $\beta $11 partially unfold is stable enough to be detected. However, the lifetimes
of the late stage intermediates are likely to be too short for experimental detection. In the subdominant unfolding route the barrel flattens after
the rupture of $\beta $11 thus exposing in excess of 50$\%$ of hydrophobic residues. As a result, we predict that there are only two detectable intermediates.

\subsubsection*{GFP Refolding Upon Force Quench} The efficacy of the SOP model was further established by following refolding after quenching an applied force from
a high value. To initiate refolding, we reduced the force on the fully stretched GFP to a quench force, $f_Q = 0$. Formation of secondary structures and establishment
of a large number of tertiary contacts occurs rapidly, in about 2.5 ms \cite{Hyeon06Structure}. Subsequently, the molecule pauses
in a metastable intermediate state in which all the secondary structural elements are formed but the characteristic barrel of the native state is
absent. The transition from the metastable intermediate to the NBA, during which the barrel forms, is the rate limiting step. Native state formation
is signaled by the closure of the barrel and the accumulation of long-range contacts between H1 and the rest of the structure. Both the size
and the end-to-end distance decrease nearly continuously and it is only in the final stages where a precipitous reduction takes place. The root mean square deviation
of the intermediate from the native state is about 20 {\AA}, whereas the final refolded structure deviates by only 3 {\AA} from the native conformation.
Contact formation at the residue level shows that the interaction between $\beta $3 and $\beta $11 and between $\beta $1 and $\beta $6 are responsible
for barrel closing. The assembly of GFP appears to be hierarchical in the sense that the secondary structural elements form prior to the establishment
of tertiary interactions. The force-quench refolding of GFP suggests that large proteins are more likely to follow hierarchical assembly than small globular proteins.  
A similar hierarchical mechanism was recently found in thermal refolding of GFP using C$_\alpha$-Go models \cite{Gosavi08PNAS}.

\subsection*{From Folding to Function: Simulations Using SOP}

The potential link between large scale allosteric transitions and function is most vividly illustrated in biological nanomachines\cite{ThirumalaiARBBS01, HerbertAnnuRev08, ValeScience00}.
To fully understand the underlying mechanism of allostery it is important to dynamically monitor the structural changes that occur in the transition
from one state to another. The great utility of the SOP model is that it can be used to probe structural changes in the reaction
cycle of biological nanomachines, GroEL \cite{Hyeon06PNAS} and kinesin \cite{Hyeon07PNAS,Hyeon07PNAS2}.

\subsubsection*{Chaperonin GroEL}
\begin{figure}[ht]
\includegraphics[width=7.00in]{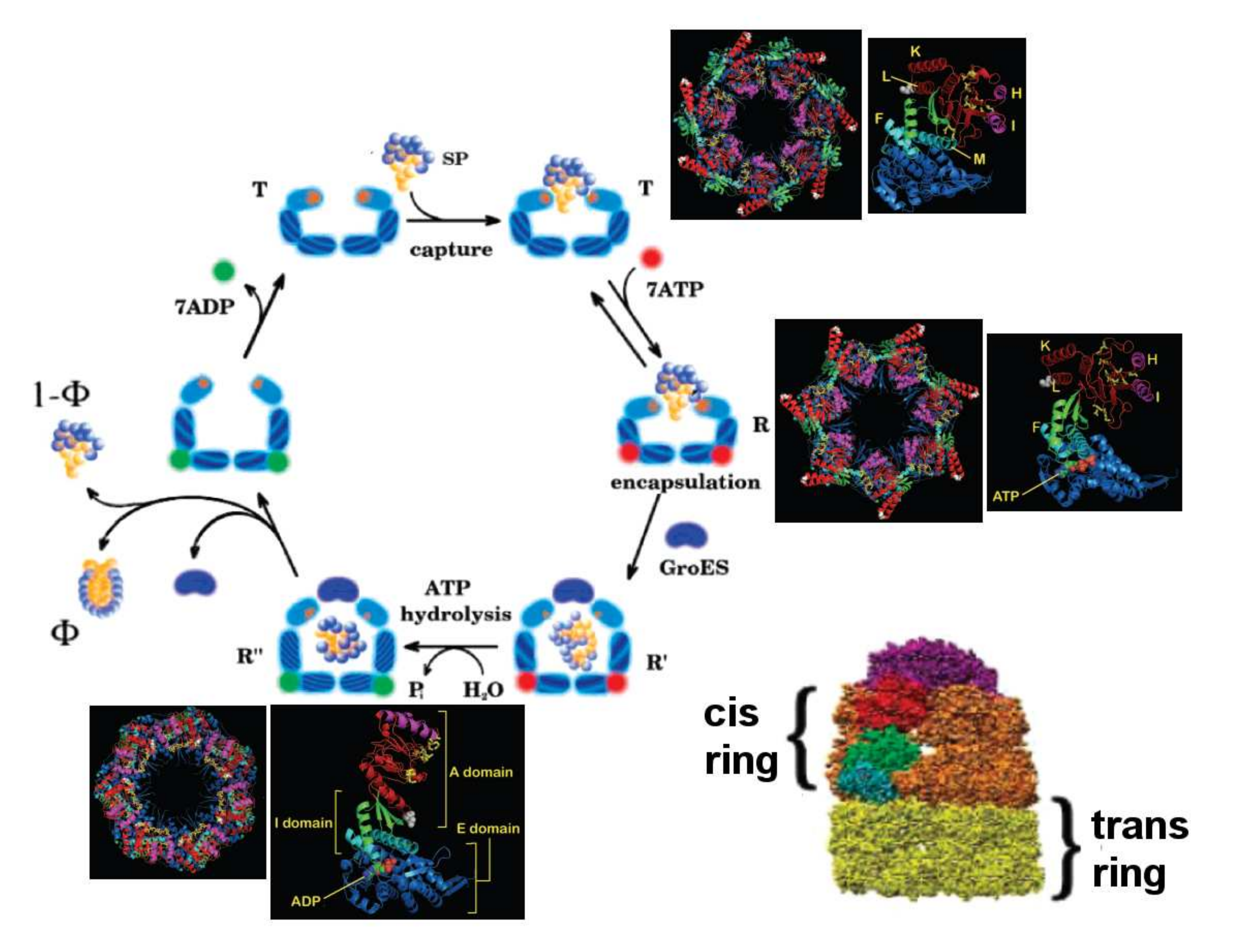}
\caption{}
\label{fig:6}
\end{figure}

The misfolding of proteins and their subsequent aggregation is linked to fatal neurodegenerative diseases like Alzheimer's and prion diseases \cite{SelkoeNature03,PrusinerPNAS98,Dobson99TBS}. 
In the cellular environment molecular chaperones, such as trigger factor \cite{Teter99Cell} or the GroEL-GroES chaperonin system powered by ATP molecules \cite{ThirumalaiARBBS01}, increase the yield of the native state for 
substrate proteins that are prone to misfold\cite{ThirumalaiARBBS01,TehverJMB08}. Thus, the normal operation of chaperonin systems are crucial to cellular function. The most well studied chaperonin is GroEL, that has two heptameric rings, stacked back-to-back. Substrate proteins are captured by GroEL in the $T$ state (Fig.\ \ref{fig:6}) while ATP-binding triggers a transition to the $R$
state. The binding of the co-chaperonin GroES requires dramatic movements in the A domains which doubles the volume of the central cavity. Although structural
and mutational studies have identified many residues that affect GroEL function, only few studies have explored the dynamics of allosteric transitions
between the various states\cite{CuiProtSci08}.

To obtain a detailed understanding of the allosteric mechanism, beyond insights gained from comparison of static structures\cite{XuNature97}, it
is important to probe the transition dynamics of the entire molecular construct. We used the SOP Hamiltonian \cite{Hyeon06Structure} to include electrostatic
interactions between charged residues and the interactions of GroEL with its ligand, ATP\cite{Hyeon06PNAS}.
{ }The order of events was monitored in the allosteric transition initiated by
ATP binding ($T \rightarrow  R$) and ATP hydrolysis ($R \rightarrow R^{\prime\prime}$). By simulating the dynamics of ligand-induced conformational
changes in the heptamer and in two adjacent subunits, we obtained an unprecedented view of the key interactions that drive the various allosteric
transitions\cite{Hyeon06PNAS}. The transitions between states are induced with the assumption that the rate of conformational changes in the molecular
machine is slower than the rate
at which ligand-binding-induced strain propagates. In the simulations, the system Hamiltonian for the GroEL molecule is switched from one pre-equilibrated
state to the other state ($T \rightarrow R$ or $R \rightarrow R^{\prime\prime}$), and the position of each interaction center is updated using
the Brownian dynamics algorithm\cite{Hyeon06PNAS, Chen07JMB}
\begin{equation}
\label{eqn:9}
\mathbf{r}_i(t+\delta t)=
\mathbf{r}_i(t)-\bm{\nabla }_{\mathbf{r}_i}
H(\{\mathbf{r}\}|X) 
\delta t/\zeta + \bm{\xi}_i(t),
\end{equation}
\noindent where the random displacement satisfies the fluctuation dissipation theorem:
\begin{equation}
\label{eqn:10}
\langle \xi_{i\alpha}(t)\xi_{i\beta}(t) \rangle = 2 \frac{k_BT}{\zeta}\delta t\delta_{\alpha \beta}\delta_{ij}, 
\end{equation}
\noindent and the system Hamiltonian for the T $\rightarrow $ 
R allosteric transition is changed from the $H(\{ \mathbf{r}\}|T)$ for pre-equilibration 
to the $H(\{ \mathbf{r}\}|R)$ for production via a switching Hamiltonian,
$H(\{ \mathbf{r}\}|T \to R)$. The changes in the Hamiltonian amount to the changes in the equilibrium distance between the
residues i and j, i.e., { }\(r_{ij}^0\) = \(r_{ij}^0(T)\), \(r_{ij}^0\) = \(r_{ij}^0(R)\) and \(r_{ij}^0\) = \(r_{ij}^0(T \to
 R)\) = \((1-f(t))r_{ij}^0(T)\) + \(f(t)r_{ij}^0(R)\) for $T$ and $R$ states and for the $T \rightarrow R$ transition.
In the implementation in Hyeon et al.\ \cite{Hyeon06PNAS} we used $f(t) = t/\tau _{TR}$.  A similar strategy that time-dependently combines two potentials of mean force has recently been
used to probe the stepping dynamics of kinesin on a microtubule \cite{Hyeon07PNAS2}.  By controlling the value of \(\tau _{TR}\), one can alter the rate of local dynamics from ATP binding
or ATP hydrolysis.
The simplicity of the SOP model allowed us to generate multiple trajectories to resolve the key events in the allosteric transitions.
Below we briefly recapitulate the major results and important testable predictions made in our preliminary study.

{\bf Heptamer dynamics show that the A domains rotate counterclockwise in the $T \rightarrow R$ transition and clockwise in $R \rightarrow 
R^{\prime\prime}$ transition}: The clockwise rotation of the apical domain alters the nature of the lining of the SP binding sites (domain color-coded in
magenta in Fig.\ \ref{fig:6}). The dynamic changes in the angle associated with the hinge motion of the intermediate (I) domain, that is perpendicular to the
A domain, lead to an expansion of the overall volume of the heptamer ring. In the $R \rightarrow R^{\prime\prime}$ transition, the A domain is erected,
so that the SP binding sites are oriented upwards to provide binding interfaces for GroES. Some residues, notably 357-361 ( Fig.\ \ref{fig:6} ), which are completely
exposed on the exterior surface in the T state, move to the interior surface during the $T \rightarrow R \rightarrow R^{\prime\prime}$ transitions.

{\bf Global $T \rightarrow R$ and $R \rightarrow R^{\prime\prime}$ transitions follow two-state kinetics}: Time-dependent changes
in root mean square deviation (RMSD) with respect to a reference state ($T$, $R$, or $R^{\prime\prime}$), differ from molecule to molecule, 
suggestive of large heterogeneity. GroEL spends a substantial fraction of time (measured in terms of first passage time) in the 
transition state (TS) region during the $T \rightarrow
R$ transition. The ensemble average of the time-dependence of RMSD for both the $T \rightarrow R$ and $R \rightarrow R^{\prime\prime}$ transitions
follow single exponential kinetics. Despite a broad transition region, the allosteric transitions can be approximately described by a two-state model.  
Interestingly, during the allosteric transitions certain regions partially unfold (i.e., GroEL behaves as a soft machine that responds to external loads).  The plastic motions, which
are indicative of malleability of GroEL, are expected to be a fundamental characteristic of all biological machines.

{\bf $T \rightarrow R$ transition is triggered by a downward tilt of helices F and M in the I-domain followed by a multiple salt-bridge
switching mechanism}: Several residues in helices F (141-151) and M (386-409) in the I domain interact with the nucleotide-binding sites in
the equatorial (E) domain thus creating a tight nucleotide binding pocket.  Tilting of the F and 
M helices by $\approx $ 15${}^{\circ}$ (Fig.\ \ref{fig:6}) enables the favorable interactions to occur. 
The $T \rightarrow R$ transition involves the formation and breakage of intra- and intersubunit contacts.  The approximate 
order of events that drive the ATP-driven $T \rightarrow R$ transition are the following (Fig.\ \ref{fig:6}): (1) The ATP-binding-induced downward tilt of the F, M helices is the earliest event that accompanies the subsequent
spectacular movement of GroEL. Upon the downward tilt of the F and M helices, the entrance to the ATP binding pocket gets narrow. In the $T$ state
E386, located at the tip of M helix, forms intersubunit salt-bridges with R284, R285, and R197. In the transition to the R state, these salt-bridges
are disrupted and a new intrasubunit salt-bridge with K80 forms simultaneously. The tilting of M helix must precede the formation of intersubunit
salt-bridge between the charged residues E386 with K80. (2) At the residue level, the reversible formation and breaking of D83-K327 salt-bridge, in concert with the intersubunit
salt-bridge switch associated with E386 and E257, are among the most significant events that dominate the $T \rightarrow R$ transition.

The coordinated global motion is orchestrated by a \emph{multiple salt-bridge switching mechanism}, and partial unfolding and stretching of 
elements in the apical domain. The movement of the A domain results
in the dispersion of the SP binding sites and also leads to the rupture of the E257-R268 intersubunit salt-bridge.  
To maintain the stable configuration in the R state, E257 engages in salt-bridge formation with
positively charged residues that are initially buried at the interface of interapical domain in the $T$ state. During the $T \rightarrow R$ transitions
E257 interacts partially with K245, K321, and R322 as evidenced by the decrease in their distances. The distance between E409-R501 salt-bridge remains
constant ($\sim $10 {\AA}) throughout the whole allosteric transitions. This salt-bridge and two others (E408-K498 and E409-K498) might be important
for enhancing positive intra-ring cooperativity and for stability of the chaperonins. In summary, coordinated dynamic changes in the network of salt-bridges 
drive the $T \rightarrow R$ transition.

{\bf $R \rightarrow R^{\prime\prime}$ transition involves a spectacular outside-in movement of K and L helices accompanied by interdomain salt-bridge
formation K80-D359}: The dynamics of the irreversible $R \rightarrow R^{\prime\prime}$ transition is propelled by substantial movements in the A
domain helices K and L.  These drive the dramatic conformational change in GroEL and result in doubling of the volume of the cavity.  
(1) Upon ATP hydrolysis the F and M helices rapidly tilt by an additional 10${}^{\circ}$. Nearly simultaneously there
is a small reduction in the P33-N153 distance \cite{Hyeon06PNAS}. These relatively small changes are the initial events in the $R \rightarrow R^{\prime\prime}$ transition.  
(2) In the subsequent step, the A domain undergoes significant conformational changes that are most vividly captured by the
outside-in concerted movement of helices K and L.{  }In the process, a number of largely polar and charged residues that are exposed to the exterior
in the $T$ state line the inside of the cavity in the $R^{\prime\prime}$ state. The outside-in motion of the K and L helices (Fig.\ \ref{fig:6}) leads to the formation of an
interdomain salt-bridge K80-D359.  These spectacular changes alter the microenvironment of the cavity interior for the substrate protein (SP).  The interaction between 
the SP and GroEL changes from being hydrophobic in the $T$ state to being hydrophilic in the $R^{\prime\prime}$ state.

The clockwise rotation of the apical domain, which is triggered by a network of salt-bridges as well as interactions between hydrophobic
residues at the interface of subunits, orients it in the upward direction so as to permit the binding of the mobile loop of GroES. Hydrophobic interactions
between SP binding sites and GroES drive the $R \rightarrow R^{\prime\prime}$ transition. The hydrophilic residues, that are hidden on the side
of apical domain in the $T$ or the $R$ state, now form an interior surface of GroEL (see the residue colored in yellow on the A domain
in Figure \ref{fig:6}).

{\bf TSEs are broad}: Disorder in the TSE structures is largely localized in the A domain which shows that the substructures in this domain
partially unfold as the barrier crossings occur (Fig.\ 6 in Hyeon et al.\ \cite{Hyeon06PNAS}). By comparison the E domain remains more or less structurally intact even at the transition state
which suggests that the relative immobility of this domain is crucial to the function of this biological nanomachine. The dispersions in the TSE
are also reflected in the heterogeneity of the distances between various salt-bridges in the transition states. The values of the contact distances,
in the $T \rightarrow R$ transition among the residues involved in the salt-bridge switching between K80, R197, and E386 at the TS have a very broad
distribution which also shows that the R197-E386 is at least partially disrupted in the TS and that K80-E386 is partially formed.

As summarized above, we probed the allosteric transitions in GroEL ($\approx 3700$ residues) using the SOP model, and produced a number 
of new predictions that can be tested experimentally. The transitions occur by a coordinated switch between networks of multiple salt-bridges. 
The most dramatic outside-in movement, the rearrangement of helices K and L of the A domain, occurs largely in the $R \rightarrow R^{\prime\prime}$ 
transition and results in intersubunit K80-D359 salt-bridge formation. In both transitions most of the conformational changes occur in 
the A domain with the E domain serving as a largely structurally static base that is needed for force transmission.  These large scale
conformational changes, which are difficult to capture using standard MD simulations, are intimately linked to function.

\subsubsection*{Kinesin}

The study of unidirectional motility of kinesin motors began with the discovery in 1985 of the kinesin{'}s ATPase activity coupled
to the unidirectional transport motion of cellular organelles along microtubules (MTs) \cite{Brady85Nature,Vale85Cell}.  The structural studies using X-ray crystallography 
\cite{Mandelkow97BC,Mandelkow97Cell,Kikkawa00Cell} and
cryo-EM \cite{ValeNature99,Kikkawa06EMBOJ} structures show that the kinesin motor has two heavy chains and two light chains.   
The heavy chain has a globular head (the motor domain)
connected via a short, flexible neck linker to the stalk, which is a long, coiled-coil region that ends in a tail region formed with a light-chain.
Single molecule experiments using optical tweezers \cite{Schnitzer97Nature,Visscher99Science,Block03PNAS} and fluorescence dye \cite{Block03Science,Yildiz04Science} 
suggested that kinesin undergoes structural transitions resulting in an alternative
binding of motor head to the microtubule binding sites that are 8-nm apart. The force-ATP-velocity (or force-ATP-randomness) relationship measured
through the single molecule assays and kinetic ensemble experiments prompted several groups to decipher the energy landscape of motor dynamics by
proposing and solving the phenomenological models that best describe the motility data \cite{Fisher01PNAS,Fisher05PNAS,Kolomeisky07ARPC,Liepelt07PRL}.  
However, understanding the working principle of kinesin motors
based on the structural changes during the reaction cycle has been missing in the study of molecular motors. Despite the rapid improvement made 
in experimental spatial and temporal resolution,
the level of observations on the kinesin dynamics using the present single molecule experiments alone is too crude to make final conclusions. In
conjunction with the experiments, we should be able to further benefit from the structure-based approach \cite{Hyeon07PNAS,Hyeon07PNAS2}.

\begin{figure}[ht]
\includegraphics[width=7.00in]{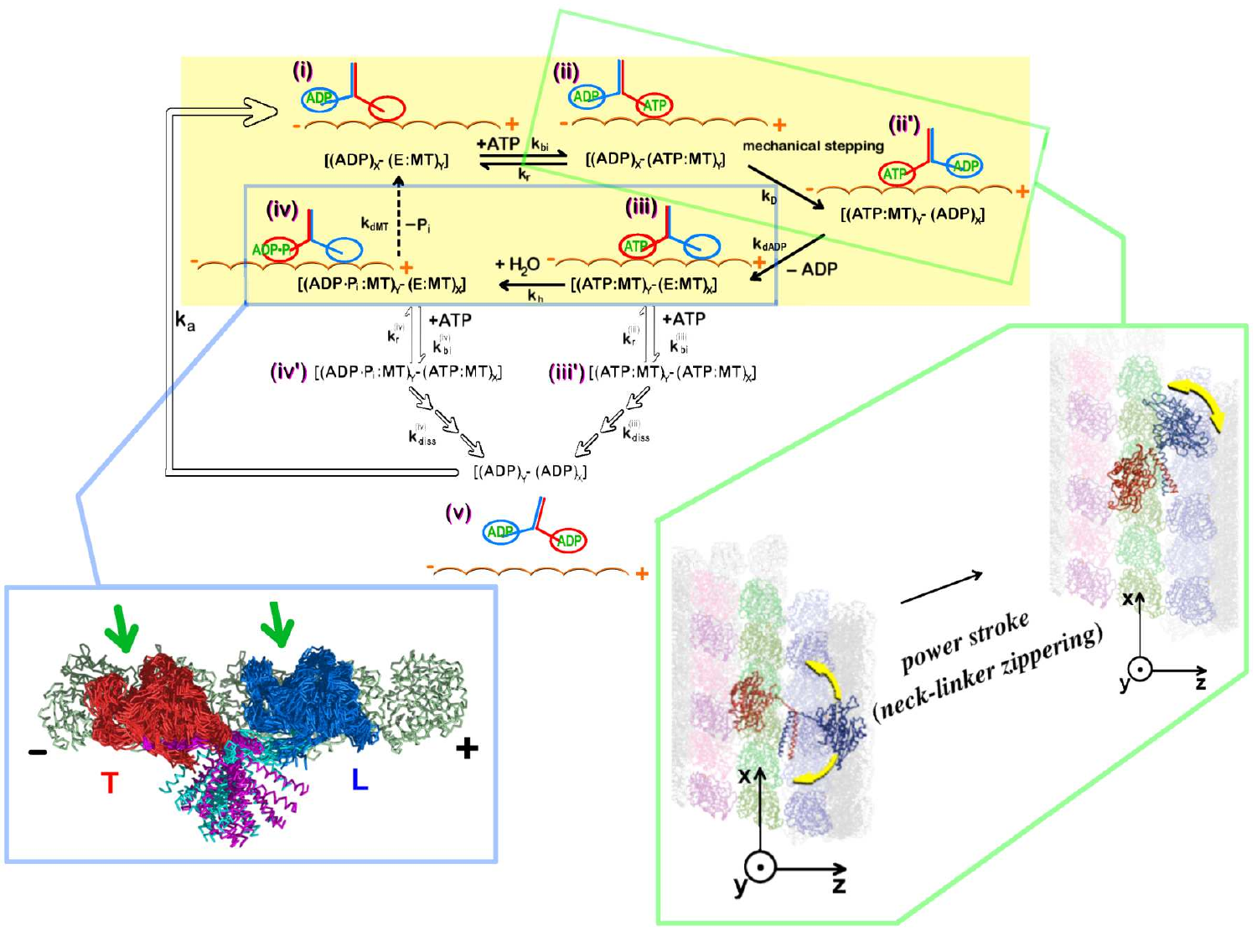}
\caption{}
\label{fig:7}
\end{figure}

In a recent study Hyeon and Onuchic (HO) \cite{Hyeon07PNAS} used the SOP model to understand the mechanochemistry of kinesin motors from a structural perspective. Treating
the MT surface as a template for the interaction between the kinesin and MT, they showed that
the topological constraint exclusively
perturbs the ATP binding pocket of the leading head through the neck-linker when both heads of the kinesin motor are bound to the microtubule binding site. The internal tension exerted through the neck-linker deforms the nucleotide
binding pocket from its native-like configuration (see structures in blue box in Fig. \ref{fig:7}). Assuming that the binding affinity of the nucleotide
to the binding pocket is maximized at the native-like configuration, the nucleotide binding to the leading head becomes chemically unfavorable.  Unless
the release of inorganic phosphate ($\text{P}_\text{i}$), leading to the dissociation of the trailing kinesin head from the microtubule binding site alleviates
the deformation of leading head structure, the ATP binding pocket of leading head remains disrupted. Therefore, the high level of processivity, unique
to the kinesin-1 motor, is achieved through the asymmetric strain induced regulation mechanism \cite{BlockPNAS06,Uemura03NSB} between the two motor domains on the MT. Computational
study using the simple structure based model clarifies the experimental proposal of the rearward strain regulation mechanism between
the two motor heads.

The above model can be extended to study the dynamic behavior of kinesin{'}s stepping motion coupled to the geometry of MT surface
(Figure \ref{fig:7}). By exhaustively sampling the configurations of kinesin tethered head on the surface of 13-protofilament MT by either modeling the neck-linker
of the MT-bound head being ordered or being disordered, HO \cite{Hyeon07PNAS2} constructed the two extreme cases of three-dimensional potentials of mean force (PMFs)
felt by the tethered head. The power stroke of the kinesin motor was mimicked by switching the PMF from the one with a disordered (unzipped) neck-linker
to the other with an ordered (zipped) neck-linker, and the stepping dynamics of kinesin tethered head was simulated using a diffusion dynamics of
a quasi-particle on the time-varying PMF. If the rate of power stroke is slower than $k_p \approx $ (20 $\mu $s)$^{-1}$, the substep of kinesin stepping
lends itself in the averaged time trace because of the sideway binding site of the MT. With an emphasis on the explicit MT topology in studying the
kinesin dynamics, this work demonstrated the interplay between the emergence of substep and the rate of power stroke. It was also shown that the
binding dynamics of kinesin to the MT is eased by a partial unfolding of kinesin structure.

The two recent applications of the SOP model to the function of biological machines \cite{Hyeon06PNAS,Hyeon07PNAS,Hyeon07PNAS2} show the utility of $C_{\alpha}$ simulations in elucidating
dynamics features that are difficult to tease out experimentally.  Furthermore, treatment of such large systems holds promise for providing
detailed (albeit at a coarse-grained level) structural perspectives in these and related ATP-consuming machines.

\section*{RNA Folding}
\begin{figure}[ht]
\includegraphics[width=5.00in]{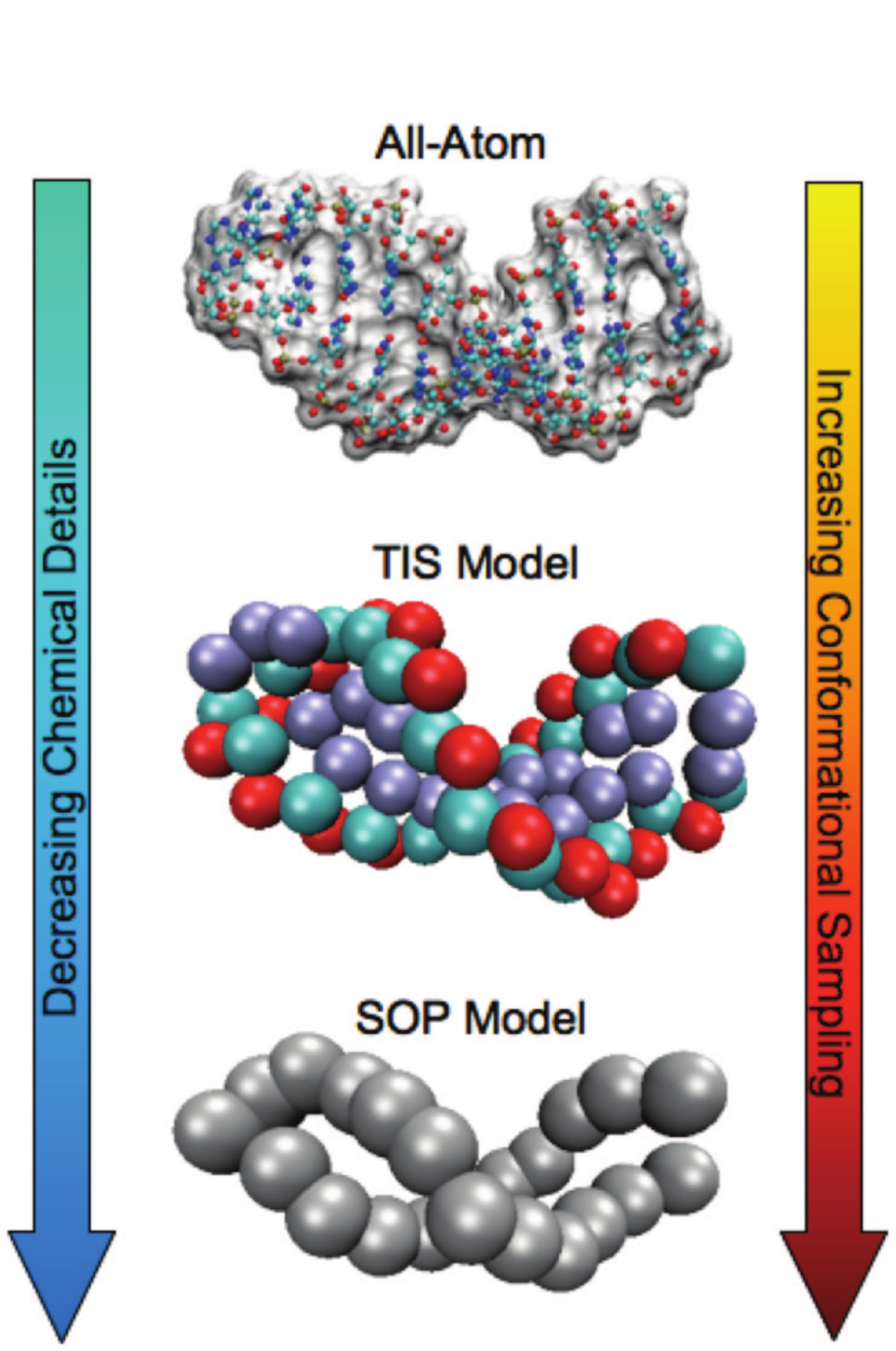}
\caption{}
\label{fig:8}
\end{figure}

Folded RNA molecules have a complex architectural organization \cite{Tinoco99JMB}. Many, not all, of the nucleotides engage in Watson-Crick 
base pairing\cite{DimaJMB05}, while other regions form bulges, loops, etc.  These structural motifs form tertiary interactions, and they give rise to a number of distinct 
folds whose stability can be dramatically altered by counterions\cite{WoodsonJMBII01}.
At first glance it might appear that it is difficult to develop 
coarse-grained models for RNA, which are polyelectrolytes, that 
fold into compact structures as the electrostatic interactions are 
attenuated by adding counterions.  Moreover, recent studies have shown valence, size, and shape of counterions profoundly influence RNA folding \cite{KoculiJMB04,Koculi07JACS,WoodsonJMBI01,WoodsonJMBII01,PanPNAS99}.  
Despite the complexity, it is possible to devise physics-based models that capture the essential aspects 
of RNA folding and dynamics.  In order to provide a framework for understanding and anticipating the outcomes of increasingly sophisticated experiments involving 
RNA we have developed two classes of models. These models are particularly useful in probing the effect of mechanical force in modulating the folding landscape of simple hairpins to ribozymes.  
In the following sections, we discuss
two coarse-graining strategies for representing RNA molecules (Fig.\ \ref{fig:8}) and assess their usefulness in reproducing experimental observations.

\subsection*{Three Interaction Site (TIS) Model\cite{HyeonPNAS05}}
From the general architecture of RNA molecules, it is immediately clear that they are composed of a series of nucleotides that are connected
together via chemically identical ribose sugars and charged phosphates that make up its backbone. { }Protruding from the backbone are four possible
aromatic bases that may form hydrogen bonding interactions with other bases, typically following the well-known Watson-Crick pairing rules.
{ }Local base-stacking interactions may also play an important role in stabilizing the folded structure.  Taking into account the above mentioned cursory observations, 
we constructed a coarse-grained off-lattice model of RNA by representing each nucleotide
by three beads with interaction sites corresponding to the ribose sugar group, the phosphate group, and the base. { }In the TIS model, the bases 
are covalently linked to the ribose center, and the sugar and phosphates make up the backbone. { }Therefore, an RNA molecule with N nucleotides is
composed of 3N interaction centers. { }The potential energy of a conformation is given by:
\begin{eqnarray}
\label{eqn:11}
\begin{split}
V_{T}&=V_{SR}+V_{LR}\\
V_{SR}&=V_{Bonds}+V_{Angles}+V_{Dihedrals}\\
V_{LR}&=V_{NC}+V_{Elec}+V_{Stack}\\
\end{split}
\end{eqnarray}
The short-range interactions ($V_{SR}$) include the bond angle, and dihedral terms
($V_{Bonds}$, $V_{Angles}$, and $V_{Dihedrals}$, respectively) which account for the chain connectivity and the angular degrees of freedom 
as is commonly used in coarse-grained models of this type \cite{KlimovFoldDes98}.  The long-range interactions ($V_{LR}$) are composed of the native interaction term,
$V_{NC}$, pairwise additive electrostatic term between the phosphates, $V_{Elec}$, and base stacking interaction term that stabilize the hairpin, $V_{Stack  }$.  
We now describe the long-range interaction terms in detail.

The native Go interaction term between the bases mimics the hydrophobicity of the purine/pyrimidine group, and a Lennard-Jones interaction between
the nonbonded interaction centers is as follows:
\begin{eqnarray}
\label{eqn:12}
V_{NC}=\sum_{i=1}^{N-1}\sum_{j=i+1}^NV_{B_iB_j}(r)+\sum_{i=1}^N\sum_{m=1}^{2N-1}{'}V_{B_i(SP)_m}(r)+\sum_{m=1}^{2N-4}\sum_{n=m+3}^{2N-1}V_{(SP)_m(SP)_n}(r)
\end{eqnarray}
A native contact is defined as two noncovalently bound beads provided they are within a cut-off distance $r_c (=7.0$ \AA{ }in the native structure.
Two beads that are beyond $r_c$ in the native structure are considered to be {``}nonnative{''}. Pairs of beads that are considered native have
the following potential:
\begin{eqnarray}
\label{eqn:13}
V_{\alpha,\beta}(r)=C_h\left[\left(\frac{r^o_{ij}}{r}\right)^{12}-2\left(\frac{r^o_{ij}}{r}\right)^6\right]
\end{eqnarray}
For beads that are nonnative, the interactions are described by:
\begin{equation}
\label{eqn:14}
V_{\alpha, \beta}(r)=C_R \left[\left(\frac{a}{r}\right)^{12}+\left(\frac{a}{r}\right)^6\right],
\end{equation}
\noindent where a = 3.4 {\AA} and $C_R = 1$ kcal/mol. The electrostatic potential between the phosphate groups is assumed to be pairwise additive:
\begin{equation}
\label{eqn:15}
V_{Elec}=\sum _{i=1}^{N-1} \sum _{j=i+1}^N {V_{P_iP_j}(r)}
\end{equation}
 { }We assume
a Debye-H{\" u}ckel interaction, which accounts for screening by condensed counterions and hydration effects, and it is given by:
\begin{equation}
\label{eqn:16}
V_{P_iP_j}=\frac{z_{P_i}z_{P_j}e^2}{4\pi\epsilon_0\epsilon_r r}e^{-r/l_D}
\end{equation}
\noindent where $z_{P_i}= -1 $ is the charge on the phosphate ion, $\ell _D$ the Debye length,
$\ell _D=\sqrt{\varepsilon _rk_BT/8\text{$\pi $k}_{\text{elec}}e^2I}$ with $k_{elec} = 8.99 \times 10^9 JmC^{-2}$ and $\varepsilon _r=10$.
To calculate the ionic strength, $I=1/2 \sum _iz_i^2c_i$, the concentration of the ions, $c_i$, is used.  Since the Debye 
screening length $\sim  \sqrt{T}$, the strength of the electrostatic interaction between the phosphate groups is temperature-dependent, even
when we ignore the variations of $\varepsilon$ with T.  At room temperature ($T \sim 300$ K), the electrostatic repulsion $V_{P_iP_j} \sim 0.5$  kcal/mol between
the phosphate groups at $r \sim 5.8$ \AA, which is the closest distance between them.  It follows that the $V_{elec}$ between phosphate groups across the base pairing
(r = 16-18 \AA) is almost negligible.

Finally, it is well known that simple RNA secondary structures are stabilized largely by stacking interactions whose context-dependent values are known (16,17).  The orientation dependent stacking interaction term is taken to be:
\begin{eqnarray}
\label{eqn:17}
V_i(\{\phi\},\{\psi\},\{r\};T)=\Delta G_i(T)&\times&e^{-\alpha_{st}\{sin^2(\phi_{1i}-\phi_{1i}^{o})+sin^2(\phi_{2i}-\phi_{2i}^{o})+sin^2(\phi_{3i}-\phi_{3i}^{o})+sin^2(\phi_{4i}-\phi_{4i}^{o})\}}\nonumber\\
&\times&e^{-\beta_{st}\{(r_{ij}-r_{1i}^{o})^2+(r_{i+1j-1}-r_{2i}^{o})^2\}}\nonumber\\
&\times&e^{-\gamma_{st}\{sin^2(\psi_{1i}-\psi_{1i}^{o})+sin^2(\psi_{2i}-\psi_{2i}^{o})\}}
\end{eqnarray}
\noindent where $\Delta G(T) = \Delta H-T\Delta S$.  The bond angles $\{\phi\}$ are $\phi_{1i} \equiv \angle S_{i}B_{i}B_{j}$, $\phi_{2i} \equiv \angle B_{i}B_{j}S_{j}$, $\phi_{3i} \equiv \angle S_{i+1}B_{i+1}B_{j-1}$, and $\phi_{4i} \equiv \angle B_{i+1}B_{j-1}S_{j-1}$.  The distance between two paired bases $r_{ij}=|B_i-B_j|$, $r_{i+1j-1}=|B_{i+1}-B_{j-1}|$, and
$\psi_{1i}$ and $\psi_{2i}$ are the dihedral angles formed by the four beads 
$B_iS_iS_{i+1}B_{i+1}$ and $B_{j-1}S_{j-1}S_jB_j$, respectively. 
The superscript $o$ refers to angles and distances in the PDB structure. 
The values of $\alpha_{st}$, $\beta_{st}$ and $\gamma_{st}$ are 
1.0, 0.3\AA$^{-2}$ and 1.0 respectively. 
The values for $\Delta H$ and $\Delta S$ were taken from Turner's thermodynamic 
data set \cite{MathewsJMB99,WalterPNAS94}.

Once the appropriate model has been formulated, simulations are performed to follow the dynamics of the RNA molecule of interest for comparison to experiments.  A combination
of forced unfolding and force quench refolding of a number of RNA molecules has been used to map the energy landscape of RNA.  These experiments identify
kinetic barriers and the nature of intermediates by using mechanical unfolding or refolding trajectories that monitor end-to-end distance $R(t)$ of the molecule in real time (t)
or from force-extension curves (FEC's).  The power of simulations is that they can be used to deduce structural details of the intermediates that cannot be unambiguously 
inferred using $R(t)$ or FEC's.  As such, forced-unfolding simulations are performed by applying a constant force to 
the bead at one end of the molecule under conditions that mimic the experimental conditions as closely as possible.  We can then 
observe their dynamics in simulations to understand the microscopic view of how they behave.

\subsubsection*{Forced Unfolding of P5GA Using the TIS Model}
To date laser optical tweezer experiments have used $f$ to unfold or refold by force-quench by keeping $T$ fixed \cite{OnoaCOSB04}.  A fuller understanding of RNA folding landscape
can be achieved by varying $T$ and $f$.  Calculations using the TIS model for even a simple hairpin show that the phase diagram is rich when both $T$ and $f$ are
varied.  Using the fraction of native contacts, $\langle Q \rangle$, as an order parameter, the diagram of states in the $(f,T)$ plane shows that the P5GA hairpin behaves approximately as a `two-state' folder.  In the absence of force $f = 0$ pN, the 
folding unfolding transition midpoint is at $T_m = 341$ K.  As force increases, $T_F$ decreases monotonically such that the transition midpoints $(T_m,f_m)$ form a phase
boundary separating the folded ($\langle Q \rangle > 0.5$ and $\langle R \rangle < 3$ nm) and unfolded states.
The phase boundary is sharp at low $T_m$ and large $f_m$, but it is broad at low force.  The locus of points separating the unfolded and folded states is given by:
\begin{equation}
\label{eqn:18}
f_c\sim f_o\left(1-\left(\frac{T}{T_m}\right)^{\alpha}\right)
\end{equation}
\noindent where $f_0$ the critical force at low temperatures and $\alpha (=6.4)$ is a sequence-dependent exponent.  The large value of $\alpha$ suggests
a weak first-order transition.

\begin{figure}[ht]
\includegraphics[width=5.00in]{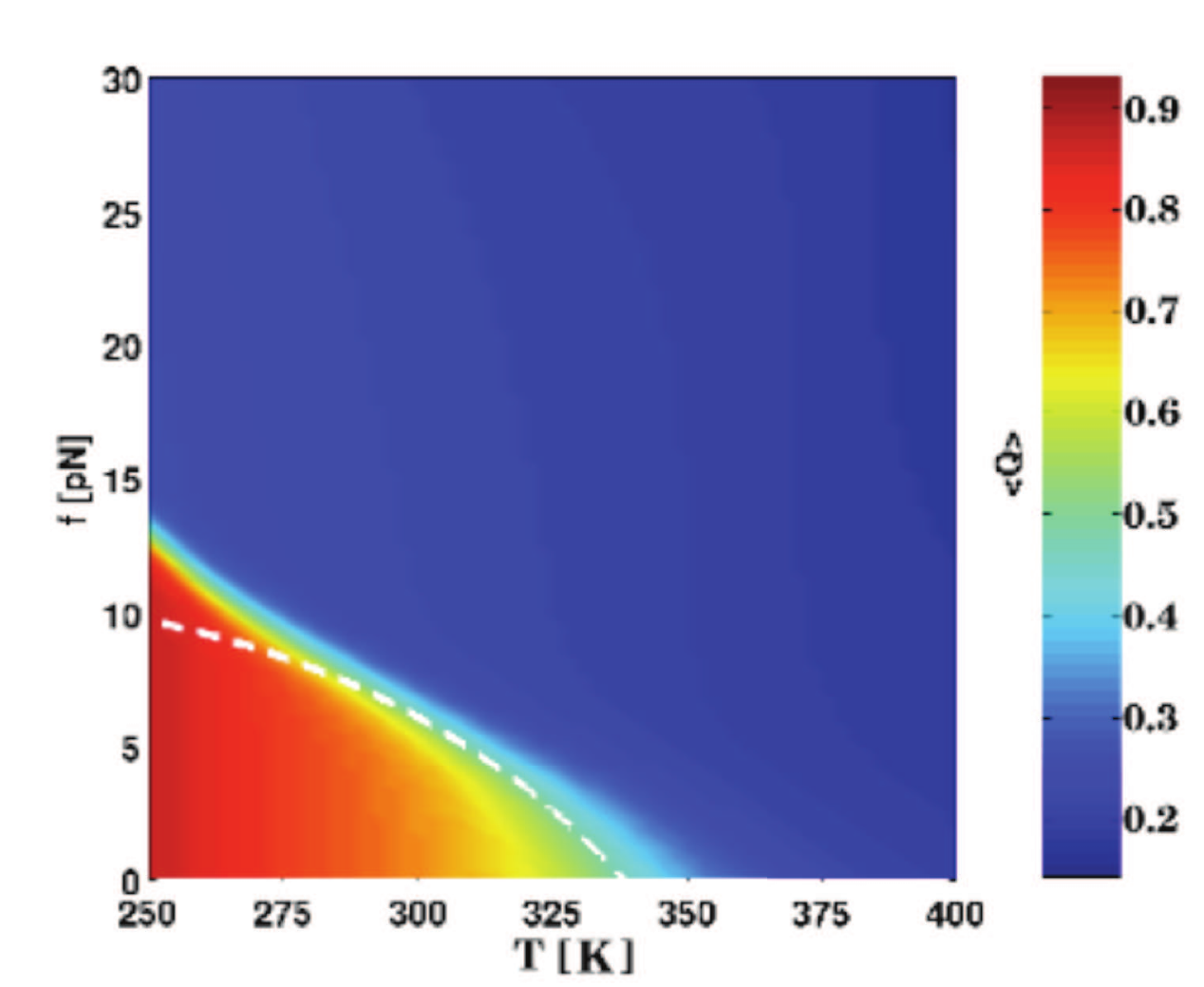}
\caption{}
\label{fig:9}
\end{figure}

\begin{figure}[ht]
\includegraphics[width=7.00in]{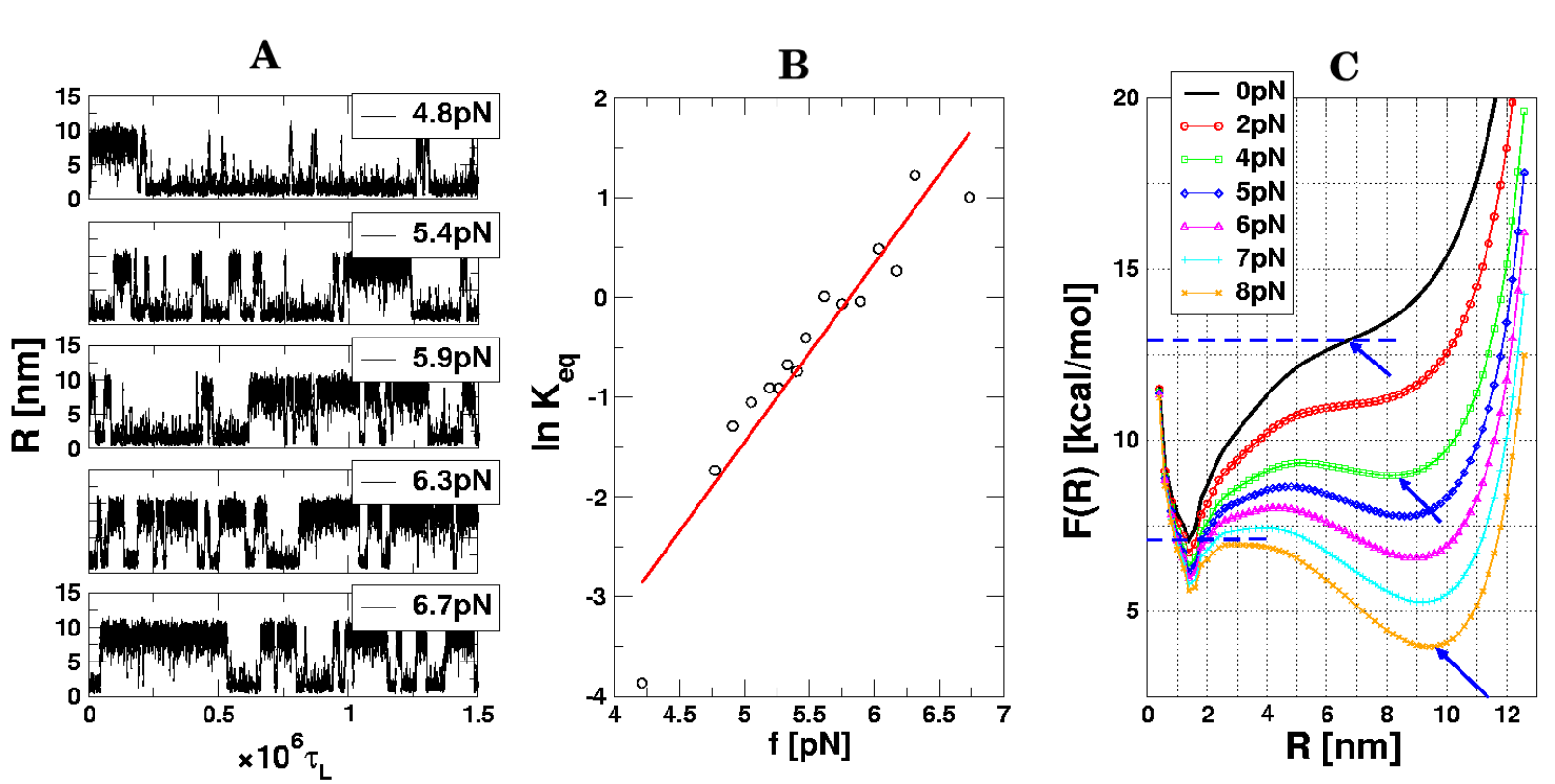}
\caption{}
\label{fig:10}
\end{figure}

The thermodynamic relation $\log K_{eq}(f) = \Delta F_{UF}/k_BT + f\cdot \Delta x_{UF}/k_BT$
and the dependence of $\log K_{eq}$
($K_{eq}$ is computed as time averages of the traces in Fig.\ \ref{fig:10}{\bf A}) on
$f$ is used to estimate $\Delta F_{UF}$ and $\Delta x_{UF}$,
which is the equilibrium distance separating the native basin of attraction
(NBA) and the basin corresponding to the ensemble of unfolded states
(UBA). The transition midpoint $K(f_m) = 1$ gives $f_m \approx 6$ pN, which is in excellent agreement with the value obtained from
the equilibrium phase diagram (Fig.\ \ref{fig:9}). From the slope,
$\partial \log K_{eq}(f)/\partial f = 1.79$ pN$^{-1}$, $\Delta x_{UF} \approx 7.5$ nm, we found, by
extrapolation to $f = 0$, that $\Delta F_{UF} \approx  6.2$ kcal/mol under the assumption
that $\Delta x_{UF}$ is constant and independent of $f$.

In the RNA pulling experiments\cite{Bustamante2},the time interval between the hopping transitions from folded to unfolded
states at the midpoint of force was measured at a single temperature. We calculated the dynamics along the phase
boundary $(T_m,f_m)$ to evaluate the variations in the free-energy
profiles and the dynamics of transition from the NBA to UBA. Along the boundary $(T_m,f_m)$,
there are substantial changes in the free-energy landscape. The
free-energy barrier $\Delta F^\ddagger$ increases dramatically at low $T$ and high
$f$.  The weakly first-order phase transition at $T \approx  T_m$ and low $f$ becomes increasingly strong 
as we move along the $(T_m,f_m)$ boundary to low $T$ and high $f$.

The two basins of attraction (NBA and UBA) are separated by a free-energy
barrier whose height increases as force increases (or temperature decreases)
along $(T_m,f_m)$. 
The hopping time $\tau_h$ along ($T_m$,$f_m$) is 
\begin{equation}
\label{eqn:20}
\tau_h=\tau_0\exp{(\Delta F^{\ddagger}/k_BT)}. 
\end{equation}
To estimate the variations in $\tau _h$ along the $(T_m,f_m)$ boundary, we performed three very long overdamped
Langevin simulations at $T_m = 305$ K and $f_m = 6$ pN. The unfolding/refolding time is observed to be
1-4 ms. From the free-energy profile, we find $\Delta F^{\ddagger}/T\sim 3$, 
so that $\tau _0 = 0.05$ to $0.2$ ms. Consequently, $\tau _h$ at $T = 254$ K and $f = 12$ pN 
is estimated to be 1-4 s, which is three orders of magnitude greater than at the higher $T_m$ and lower $f_m$.  These simulations showed that only by probing the dynamics over a wide range of $(T,f)$ values can the entire energy landscape be constructed.

To probe the structural transitions in
the hairpin, we performed Brownian dynamics simulations at a constant force with
$T = 254$ K. From the phase diagram, the equilibrium unfolding force at this temperature is $12$ pN (Fig.\ \ref{fig:9}). 
To monitor the complete unfolding of P5GA, in the time course of the
simulations, we applied $f = 42$ pN to one end of the hairpin with the other
end fixed. In contrast to thermal unfolding (or refolding), the initially closed hairpin unzips from the end to the loop region.
The unzipping dynamics, monitored by the time dependence of R, shows `quantized staircase-like
jumps' with substantial variations in step length, that depend on the initial conditions. The lifetimes associated with the
`intermediates' vary greatly. The large dispersion reflects the heterogeneity of the mechanical unfolding pathways. 
Approach to the stretched state that occurs in a stepwise `quantized' manner' \cite{KlimovPNAS99}, as was first shown in lattice
models of proteins \cite{KlimovPNAS99}.

\subsubsection*{Force-Quench Refolding \cite{HyeonPNAS05}}
To monitor the dynamics of approach to
the NBA, we initiated refolding from extended conformations with $R = 13.5$ nm,
prepared by stretching at $T = 290$ K and $f = 90$ pN. Subsequently, we quenched the force to $f = 0$, and
the approach to the native state was monitored. From the distribution
of first passage times, the refolding kinetics follow exponential kinetics with the mean folding time
of $\approx 191$ $\mu s$, compared with $12.4$ $\mu s$ in the temperature quench. It
is remarkable that, even though the final conditions ($T = 290$ K and $f = 0$) are the same as in
thermal refolding, the time scale for hairpin formation upon force quench is significantly large than thermal refolding.

The large difference arises because the molecules that are fully stretched with $f \gg f_m$ and those that are 
generated at high $T$ have vastly different initial conformations.  Hence, they can
navigate entirely different regions of
the energy landscape in the approach to the native conformation. The distribution of $R$ in the thermally denatured conformations
is $P(R) \propto e^{-\beta V_{tot}(R)/k_BT_0}$ ($T_0$ is the initial temperature), 
whereas in the ensemble of the stretched conformation have $P(R)\propto \delta(R-R_s)$ where $R_s$ is the value of R when the hairpin is fully extended. The initially stretched
conformations ($R _{ext} = 13.5$ nm) do not overlap with the accessible 
regions of the canonical ensemble of thermally denatured conformations \cite{HyeonBJ06}.
As a consequence, the regions of the free-energy landscape from which folding commences in force-jump folding are
vastly different from those corresponding to the initial population of thermally equilibrated ensemble.

The pathways explored by
the hairpins en route to the NBA are heterogeneous. Different
molecules reach the hairpin conformation by vastly different routes. Nevertheless, the time dependence of $R$ shows that
the approach to the native conformation occurs in stages. Upon release of force, 
there is a rapid initial decrease in $R$ that results in the collapse
of the hairpin. Surprisingly, this process takes an average of several microseconds, 
which is much longer than expectations based on theories of collapse
kinetics of polymer coils \cite{ThirumJPI, PitardEL98}. In the second stage, the hairpin fluctuates in relatively compact state with $R$ in the broad
range (25-75 \AA) for prolonged time periods. On these time scales, which vary considerably depending on the molecules, conformational search occurs
among compact structures. The final stage is characterized by a further decrease in $R$ that 
takes the molecules to the NBA. The last stage is the most cooperative
and abrupt, whereas the first two stages appear to be much more
continuous. Interestingly, similar relaxation patterns characterized by heterogeneous pathways and continuous
collapse in the early stages have been observed in force-quench refolding of ubiquitin \cite{FernandezSCI04}. 
The multistage approach to the native stage is reminiscent of the three-stage refolding by Camacho-Thirumalai for protein refolding \cite{CamachoPNAS93}.

\subsection*{SOP Model for RNA Folding}
The TIS interaction model is not the simplest possible representation
of RNA molecules, and one can further simplify the representation of RNA when the number of nucleotides is large.  Instead
of representing each nucleotide by three beads, like the protein counterparts, 
we can represent each nucleotide by a single bead.  Such a model is similar to the SOP representation 
of proteins.  The interactions stabilizing the native conformation are taken to be uniform.  However, variations of
this model are required for accurate modeling of RNA structures that have a subtle interplay between secondary and tertiary
interactions.  

One of the computational bottlenecks of MD simulations is the computation
of the torsion angle potential, largely because of the calculation of the trigonometric function in the
energy function.  The repeated calculation of the dihedral angle potential term is sufficiently burdensome that some choose to use look-up tables 
so that its calculation are done only at the beginning of the program run.  If the configuration of the torsion angle potential is not required then in simulation efficiency, an appreciable increase would be achieved, making such an approach attractive if it is reasonable.  These arguments were the basis for the construction of the SOP model.  In this very simple model, a single bead represents each nucleotide.  
Local interactions are defined by bond potentials and native contacts determine favorable long-range interactions.  
The Hamiltonian for the SOP model is the same as for proteins except the values of the parameters are different (see Table 1 in Hyeon et al.\ \cite{Hyeon06Structure}).

\subsubsection*{Stretching Azoarcus Ribozyme}
\begin{figure}[ht]
\includegraphics[width=5.00in]{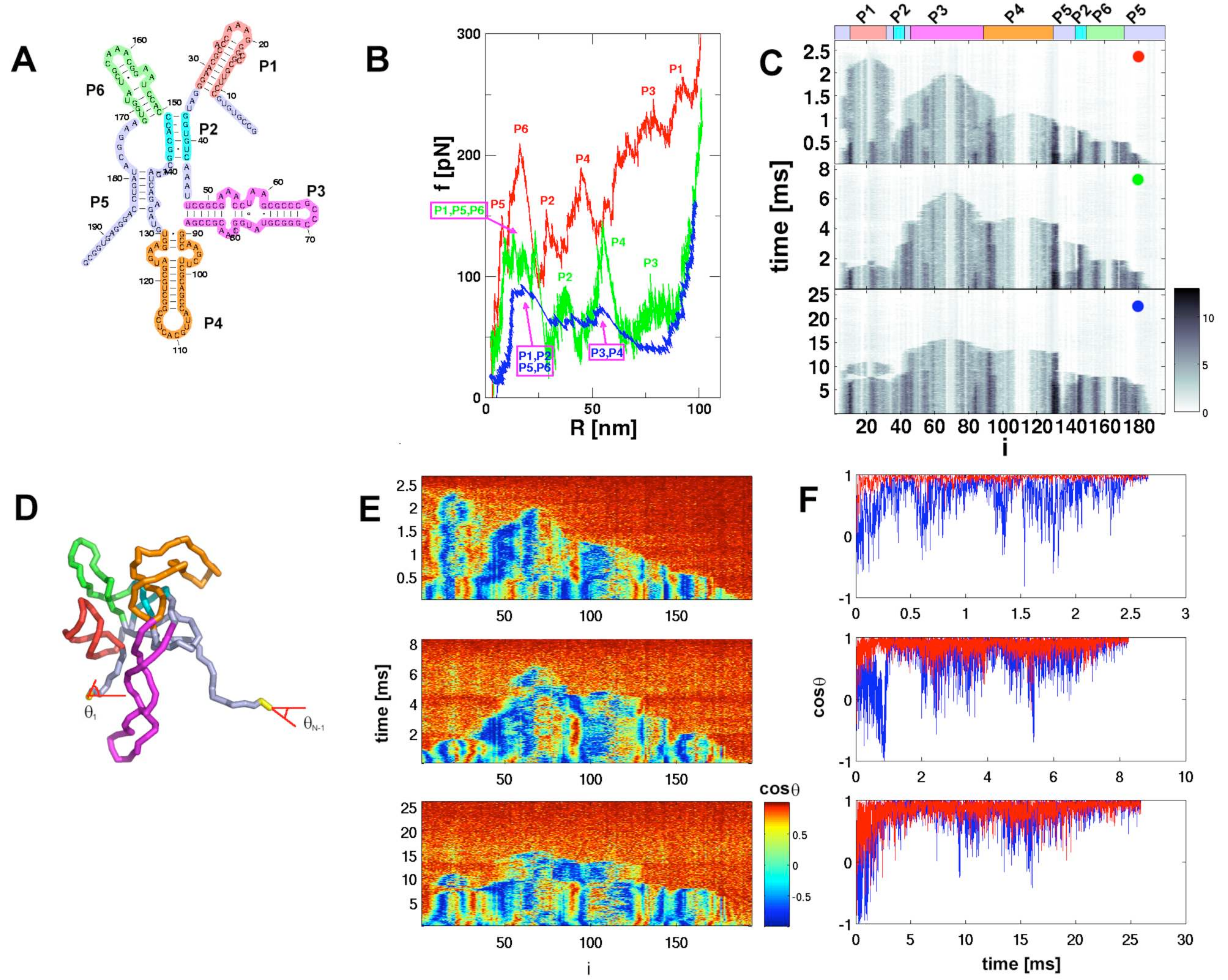}
\caption{}
\label{fig:14}
\end{figure}

SOP model simulations of the rip dynamics of the \textit{Azoarcus} ribozyme were
performed (Figure \ref{fig:14}{\bf A}). The structure of the (195 nt) \textit{Azoarcus} ribozyme \cite{RanganPNAS100} (PDB code: 1u6b) 
is similar to the catalytic core of the \textit{T.\ thermophila} ribozyme,
including the presence of a pseudoknot. The size of this system in terms
of the number of nucleotides allows exploration of the forced unfolding
over a wide range of loading conditions.

For the \textit{Azoarcus} ribozyme, ten mechanical unfolding trajectories were generated
at three loading rates. At the highest loading rate, the FEC has
six conspicuous rips (red FEC in Figure \ref{fig:14}{\bf B}), whereas at the lower $r_f$ the number of peaks is reduced to between
two and four. The structures in each rip were identified by comparing the FEC's (Figure \ref{fig:14}{\bf B}) with the 
history of rupture of contacts (Figure \ref{fig:14}{\bf C}). At the
highest loading rate, the dominant unfolding pathway of the \textit{Azoarcus} ribozyme is $N \rightarrow [P5] 
\rightarrow [P6] \rightarrow [P2] \rightarrow [P4] \rightarrow [P3] \rightarrow [P1]$. At
medium loading rates, the ribozyme unfolds via $N \rightarrow [P1, P5, P6] \rightarrow [P2] \rightarrow [P4] \rightarrow [P3]$, 
which leads to four rips in the FEC's. At the lowest loading rate, the number of rips is further reduced to two, 
which we identify with $N \rightarrow [P1, P2, P5, P6] \rightarrow [P3, P4]$.
Unambiguously identifying the underlying pulling speed-dependent conformational changes requires not only the FEC's, 
but also the history of rupture of contacts (Figure \ref{fig:14}{\bf C}). The simulations using the SOP model also showed that unfolding pathways can be altered by varying the loading rate.

To understand the profound changes in the unfolding pathways as $r_f$ is
varied, it is necessary to compare $r_f$ with $r_T$, 
the rate at which the applied force propagates along RNA (or proteins) \cite{Hyeon06Structure}.
In both AFM and LOT experiments, force is applied to one end of the chain (3$^\prime$ end) while the
other end is fixed. The initially applied tension propagates over time
in a nonuniform fashion through a network of interactions that stabilize the native conformation. The
variable $\lambda = r_T/r_f$ determines the rupture history of the biomolecules.
If $\lambda \gg 1$, then the applied tension at the 5$^\prime$ end of the RNA propagates rapidly
so that, even prior to the realization of the first rip, 
force along the chain is uniform. This situation pertains to the LOT experiments (low $r_f$). In the opposite limit, 
$\lambda \ll 1$, the force is nonuniformly felt along the chain. In such a situation,
unraveling of RNA begins in regions in which the value of local
force exceeds the tertiary interactions. Such an event occurs close to
the end at which the force is applied.

The intuitive arguments given above were made precise by computing the
rate of propagation of force along the \textit{Azoarcus} ribozyme. To visualize
the propagation of force, we computed the dynamics of alignment of the angles between the bond
segment vector ($\mathbf{r}_{i,i+1}$) and the force direction during the unfolding process (Figures \ref{fig:14}{\bf D}-{\bf F}). The nonuniformity in
the local segmental alignment along the force direction, which results in a heterogeneous distribution of times in which segment vectors
approximately align along the force direction, is most evident at the
highest loading rate (Figure \ref{fig:14}{\bf E}). Interestingly, the dynamics of the force propagation occurs sequentially from one end
of the chain to the other at high $r_f$. Direct comparison of the
differences in the alignment dynamics between the first ($\theta _1$) and last angles ($\theta _{N-1}$) (see Figure \ref{fig:14}{\bf D}) illustrates the discrepancy in the
force values between the $3^{\prime}$ and $5^{\prime}$ ends (Figure \ref{fig:14}{\bf F}). There is nonuniformity
in the force values at the highest $r_f$, whereas there is a more homogeneous alignment at low $r_f$. The microscopic
variations in the dynamics of tension propagation are reflected in the
rupture kinetics of tertiary contacts (Figure \ref{fig:14}{\bf C}) and, hence, in the dynamics of the rips (Figure \ref{fig:14}{\bf B}).

These results highlight an important prediction of the SOP model, that
the very nature of the unfolding pathways can drastically change depending
on the loading rate, $r_f$. The dominant unfolding rate depends on $r_f$, suggesting that
the outcomes of unfolding by LOT and AFM experiments can be dramatically different.
In addition, predictions of forced unfolding based on all-atom MD simulations should
also be treated with caution unless, for topological reasons (as in the Ig27 domain from muscle protein titin),
the unfolding pathways are robust to large variations in the loading rates.

\section*{Concluding Remarks}
We have presented a handful of applications to show the power of using
simple coarse-grained structure-based models in the context of folding
and functions of RNA and proteins. At a first glance it seems
remarkable that such simple models can capture the complexity of
self-assembly and, more impressively, describe in great detail the
conformational dynamics of molecular machines.  However, theoretical
arguments and simulations of lattice models demonstrate that the
dominance of native interactions that cooperatively stabilize
the folded structures over non-native contacts (that occur more non-specifically) is the reason
for the success of the structure-based approaches.

There are several avenues that are likely to be explored using coarse
grained models of increasing sophistication. First, experiments are
starting to provide detailed information on the structures of unfolded
states of proteins in the presence of denaturants such as urea and
guanadinum hydrochloride. Direct simulations, therefore, require
models of denaturants within the context of the CG models. Preliminary
studies that tackle this challenging problem have already appeared\cite{OBrien08JACS2}.
Similarly, there is a challenge to model the counterion-dependent
nature of unfolded states of ribozymes.  This will require
incorporating in an effective way counterion size and shape within the
CG models.  Second, it is increasingly clear that functions require
interactions between biomolecules. Thus, the CG models will have to be
expanded to include scales ranging from microns (DNA) to nanometers
(RNA and proteins). Third, the brief description of the  molecular
machines given here shows a complex relationship between the mechanochemical
cycles and functions.  Explaining the linkage between the
conformational changes for biological machines will require progress
in establishing the validity of the CG models as well further
developments in refining them. These and other challenges and progress
to date show that the next ten years will witness an explosion in
routinely using CG models to quantitatively understand many phenomena
ranging from folding to function.

\section*{Acknowledgments}
We thank several previous group members, notably, Prof.\ Carlos J.\ Camacho, Prof.\ Margaret S.\ Cheung, Prof.\ Ruxandra I.\ Dima and Dr.\ J.\ D.\ Honeycutt 
for valuable contributions.  This work was supported in part by a grant from the National Science Foundation (CHE 05-14056).  DLP and SSC are each supported 
by Ruth L. Kirschstein National Research Service Awards from the National Institutes of Health.

\providecommand{\refin}[1]{\\ \textbf{Referenced in:} #1}

\newpage
\section*{Figure Captions.}

\noindent {\bf Fig.~\ref{fig:1}}: Coarse grained representation of polypeptide chains.   In a lattice model ({\bf{A}}) beads are confined to occupying the vertices of a suitable lattice, while in an off-lattice model ({\bf{B}}) beads of the chain can 
occupy any position consistent with the underlying ( typically continuous and differentiable ) Hamiltonian and equations of motion.  The schematic representation in ({\bf A}) shows a
folded structure in a cubic lattice with $N = 27$.
(Figures generate with VMD \cite{HumphreyJMG96} and Mathematica \cite{Mathematica:2007}).

\noindent {\bf Fig.~\ref{fig:2}}: The native conformation of a sequence of a 2D 23-mer lattice model to probe the role of disulfide bonds in folding.  
The sequence consisted of hydrophobic (H), polar (P), and
Cys (C) residues.  Exhaustive Monte Carlo simulations were used to examine the role of non-native intermediates in protein folding \cite{CamachoPROTEINS95}.  [2-15;9-22] form disulfide bonds in the native state (squares).

\noindent {\bf Fig.~\ref{fig:3}}: Camacho and Thirumalai \cite{CamachoPROTEINS95}  showed that there are many complex pathways leading to the native state [2-15;9-22].  The figure reveals that a
nonzero number of trajectories pass the native-like intermediates ({\bf{I}$_1$} and {\bf{I}$_2$}).  Non-native intermediates are only sampled early in the folding reaction.  
Time was measured in Monte Carlo Steps (MCS).

\noindent {\bf Fig.~\ref{fig:4}}: Schematic of rugged folding landscape of a foldable sequence.  
The potentials of mean-force illustrate a central Native Basin of Attraction (NBA) flanked by two native-like
metastable minima of slightly higher energy.  The flanking minima are separated from the central minimum by transition states ( $\ddag$ at left ).  It is important to
bear in mind that this is a simple illustration and that many foldable sequences do not get trapped in metastable minima.  Nevertheless, the concept is important in
the context of aggregation-prone proteins (e.g., PrP$^{C}$) (Figures generated with VMD \cite{HumphreyJMG96} and Mathematica \cite{Mathematica:2007}).

\noindent {\bf Fig.~\ref{fig:5}}: ({\bf{A}}) Native structure of GFP (PDB ID 1GFL) that shows the characteristic barrel structure.  ({\bf{B}}) Illustration of the connectivity of the
various secondary structure elements.  ({\bf{C}}) The force-extension curve extracted from constant-loading rate simulations at $v = 2.5 \mu$m/s and 
with a spring constant of 35 pN/nm that is typical of the values used in simulations.  ({\bf{D}})  The primary unfolding pathways extracted from the simulations; 72\% followed the dominant (top)
pathway, while 28\% followed an alternate (bottom) pathway.  The partitioning shown here for GFP has also been observed in forced-unfolding of T4-lysozyme \cite{LiPNAS08}.

\noindent {\bf Fig.~\ref{fig:6}}: The hemicycle of GroEL heptamer ({\it cis}-ring only), which is completed in about 6 secs at $37^{\circ}$ C in the presence of substrate protein and ATP.  Upon ATP binding GroEL undergoes $T \rightarrow R$ transition, while interaction with GroES and subsequent ATP hydrolysis results in $R \rightarrow R^\prime \rightarrow R^{\prime\prime}$ transitions.  X-ray structures of $T$ and $R^{\prime\prime}$ have been determined.  The $R$ structure is known from cryo-EM maps.  At each stage of the mechanochemical cycle, defined by the chemical state of nucleotide and substrate protein, the GroEL structure changes dramatically. Top views of the GroEL heptamer at T, R, and R" states are shown, and the nomenclature of domains and helices are also given in the structure of a single subunit. The full GroEL structure with double ring is shown in the right at the bottom. 

\noindent {\bf Fig.~\ref{fig:7}}: Mechanochemical cycle of conventional kinesin (kin-1). ({\bf A}) During the kinetic step shown in the blue box, ATP binding to the leading head is inhibited, which leads to the high level of processivity of the kinesin motor. This aspect is explained by the mechanochemistry due to the asymmetric strain induced regulation mechanism between the two motor domains on the microtubule (MT). The thermal ensemble of structures from the simulations shows that the nucleotide binding pocket of the leading head (L) is more disordered than that of the trailing head (T).  Both are indicated by the green arrows. The conformation of L is maintained as long as $T$ remains bound to the MT. The tension built on the neck-linker of the L leads to the disorder in the ATP binding pocket. ({\bf B}) The kinetic step from (i) to (ii) enclosed in the green box denotes the stepping dynamics of kinesin motor, which is explained by the combined processes of power stroke and diffusional search of the next binding site. Because of the multiplicity of the MT binding sites, the pattern of time traces involving stepping dynamics can be affected by the rate of power stroke. 

\noindent {\bf Fig.~\ref{fig:8}}: A schematic illustration of the various levels of coarse-graining for models of RNA.  The detailed all-atom representation (top) can be reduced to include three beads for each nucleotide corresponding to the base, sugar, and phosphate moieties as in the TIS model (center).  Further coarse-graining results in each bead being represented by a single nucleotide (bottom), and is referred to as the SOP model.  The energy functions in the TIS and SOP models are shown in Eqs. \ref{eqn:11} and \ref{eqn:8}, respectively.

\noindent {\bf Fig.~\ref{fig:9}}: 
Phase diagram for the P5GA hairpin in terms of $f$ and $T$. This panel shows the diagram of states obtained using the 
fraction of native contacts as the order parameter. 
The values of the thermal average of the fraction of native contacts, $\langle Q\rangle$, are color coded as indicated on the scale shown on the right. 
The dashed line is a fit to the locus of points in the ($f,T$) plane 
that separates the folded hairpin from the unfolded states (Eq.\ \ref{eqn:18}). 

\noindent {\bf Fig.~\ref{fig:10}}:
({\bf A})  Time traces of $R$ at various values of constant force at $T=305$ K.
At $f=4.8$ pN$ <f_m\approx 6$ pN, $\langle R\rangle$ fluctuates around at low values which shows that 
the NBA is preferentially populated (first panel).  
At $f\sim f_m$ (third panel) the hairpin hops between the folded state (low $R$ value) and unfolded states ($R\approx 10$ nm).
The transitions occur over a short time interval. 
These time traces are similar to that seen in Fig.2-C of Liphardt et al.\ \cite{Bustamante2}. 
({\bf B}) Logarithm of the equilibrium constant $K_{eq}$ (computed using the time traces in ({\bf A})) 
as a function of $f$. 
The red line is a fit with $\log{K_{eq}}=10.4+1.79 f$. 
({\bf C}) Equilibrium free energy profiles $F(R)$ as a function of $R$ at $T=305$ K.
The colors represent different $f$ values that are displayed in the inset. The arrows give the location of the unfolded basin of 
attraction.  Note that the transition state moves as a function f in accord with the Hammond postulate.

\noindent {\bf Fig.~\ref{fig:14}}:
({\bf A}) Secondary structure of \textit{Azoarcus} ribozyme. 
({\bf B}) Force-extension curves of \textit{Azoarcus} ribozyme at three $r_{f}$ ($v = 43 \mu$m/s, $k_{s} = 28 $pN/nm in red, $v = 12.9 \mu$m/s, $k_{s} = 28$ pN/nm in green, and $v = 5.4 \mu$m/s, $k_{s} = 3.5$ pN/nm in blue) obtained using the SOP model. 
({\bf C}) Contact rupture dynamics at three loading rates. The rips, resolved at the nucleotide level, are explicitly labeled. 
({\bf D}) Topology of \textit{Azoarcus} ribozyme in the SOP representation. The first and the last alignment angles between the bond-vectors and the force direction are specified. 
({\bf E}) Time evolutions of $\cos\Theta_{i} (i=1,2, ...,N-1)$ at three loading rates are shown. The values of $\cos\Theta_{i}$ are color-coded as indicated on the scale shown on the right of bottom panel. 
({\bf F}) Comparisons of the time evolution of $\cos\Theta_{i} (blue)$ and $\cos\Theta_{N-1}$ (red) at three loading rates shows that the differences in the $f_{c}$ values at the opposite ends of the ribozyme are greater as $r_{f}$ increases. 

\newpage
\section*{Figures.}

\end{document}